\documentclass[sort&compress]{elsarticle}

\usepackage[utf8]{inputenc}
\usepackage[T1]{fontenc}
\usepackage{amsmath,amssymb,amsfonts}
\usepackage{graphicx}
\graphicspath{{figures/}}
\usepackage{booktabs}
\usepackage{multirow}
\usepackage{tabularx}
\usepackage{longtable}
\usepackage{siunitx}
\usepackage{xcolor}
\usepackage{algorithm}
\usepackage{algpseudocode}
\usepackage{subcaption}    
\usepackage{hyphenat}
\usepackage{listings}
\usepackage{chngcntr} 
\usepackage{sty/homanlab}
\DeclareMathOperator{\sinc}{sinc}

\journal{arXiv}

\begin{document}

\begin{frontmatter}

\title{Coherence in the brain unfolds across separable temporal regimes}

\author[1]{Davide Staub\textsuperscript{\textdagger}}
\author[1]{Finn Rabe\textsuperscript{\textdagger}}
\author[1]{Akhil Misra}
\author[1]{Yves Pauli}
\author[1]{Roya Hüppi}
\author[2]{Ni Yang}
\author[1]{Nils Lang}
\author[1,3,7]{Lars Michels}
\author[1]{Victoria Edkins}
\author[4,5]{Sascha Frühholz}
\author[6]{Iris Sommer}
\author[2]{Wolfram Hinzen}
\author[1,7]{Philipp Homan\corref{cor1}}
\ead{philipp.homan@bli.uzh.ch}
\nonumnote{\textdagger These authors contributed equally to this work.}

\cortext[cor1]{Corresponding author.}

\address[1]{Department of Adult Psychiatry and Psychotherapy, University of Zurich, Zurich, Switzerland}
\address[2]{Department of Translation and Language Sciences, University Pompeu Fabra, Barcelona, Spain}
\address[3]{Department of Neuroradiology, Clinical Neuroscience Center, University Hospital Zurich, Zurich, Switzerland}
\address[4]{Department of Psychology, University of Oslo, Oslo, Norway}
\address[5]{Cognitive and Affective Neuroscience Unit, University of Zurich, Zurich, Switzerland}
\address[6]{Center for Clinical Neuroscience and Cognition, University of Groningen, Groningen, Netherlands}
\address[7]{Neuroscience Center Zurich, University of Zurich and ETH Zurich, Zurich, Switzerland}

\begin{abstract}
To maintain coherence in language, the brain must satisfy key competing temporal
demands: the gradual accumulation of meaning across extended context (drift) and the rapid
reconfiguration of representations at event boundaries (shift). 
How these processes are implemented in the human brain during
naturalistic listening remains unclear. Here, we tested whether both can
be captured by annotation-free drift and shift signals and whether their neural expression shows distinct regional preferences across the brain. These signals were derived from a large language model (LLM) processing the narrative input. To enable high-precision voxelwise encoding models with stable parameter estimates, we densely sampled one healthy adult
across more than 7 hours of listening to crime stories while collecting 
7 Tesla fMRI data. We then modeled the feature-informed hemodynamic response using a regularized encoding framework validated on independent stories. Drift
predictions were prevalent in default-mode network hubs, whereas shift predictions
were evident bilaterally in the primary auditory cortex and language association cortex. Together, these findings show that coherence during language comprehension is implemented through distinct but co-expressed neural regimes of slow contextual integration and rapid event-driven reconfiguration, offering a mechanistic entry point for
understanding disturbances of language coherence in psychiatric disorders.

\end{abstract}
\end{frontmatter}

\section*{Introduction}\label{sec1}
Coherence is a defining property of language and thought whereby meanings
accumulate across time, remain constrained by context, and yet must be
flexibly updated when situations change. This capacity underpins
narrative comprehension and reasoning and is profoundly disrupted in
psychopathology, most notably in formal thought disorder
\cite{Andreasen1979,Homan2014,Kircher2018,Cavelti2018,Palaniyappan2023,Surbeck2025}. Despite
its centrality, coherence remains poorly specified at the neural
level. Classical neurobiological models of language emphasize local
linguistic operations within a largely left-lateralized frontotemporal
system \cite{Ojemann1989,Fedorenko2009,Price2012}, while naturalistic
fMRI studies highlight distributed semantic representations across
association cortex
\cite{Huth2016,Simony2016,Yeshurun2017,Chang2022}. More recently, deep
neural networks have been shown to robustly predict brain responses
during continuous language comprehension
\cite{Schrimpf2021,Caucheteux2022,Goldstein2022,Antonello2024}. Yet
even within this emerging framework, coherence itself remains an
underspecified construct.

At a computational level, maintaining coherence entails at least two
distinct temporal operations. Meaning must be integrated gradually
across extended context, producing smooth representational drift over
seconds to minutes
\cite{Lerner2011,Honey2012,Simony2016,Chang2022}. At the same time,
context must be rapidly reconfigured at event boundaries, when shifts
in topic or scene render prior interpretations obsolete
\cite{Zacks2007,Kurby2008,Baldassano2017,Chen2017}. Neural correlates
of these operations appear partially dissociable, with default-mode
and temporoparietal networks exhibiting long integration windows and
sensitivity to higher-order narrative structure
\cite{Lerner2011,Simony2016,Yeshurun2017,Nguyen2019,Song2021} and
auditory and language-selective cortex showing rapid,
boundary-sensitive responses
\cite{Whitney2009,Geerligs2022,Anurova2022}. However, these temporal
regimes are typically examined in isolation, using incompatible
paradigms and analytic frameworks, leaving unresolved how gradual
integration and discrete reconfiguration jointly implement coherence
during naturalistic comprehension
\cite{DuBrow2017,Hasson2018,Nastase2020,Nastase2021}.

Progress on this problem has been constrained by the absence of
stimulus-derived signals that jointly capture continuous context
accumulation and discrete reconfiguration. Data-driven segmentation
methods applied to neural activity can identify when large-scale
reorganization occurs, but remain agnostic to the computational
signals driving these transitions
\cite{Baldassano2017,Chen2017,Hasson2018,Nastase2021}. Conversely,
encoding models have largely emphasized local linguistic structure,
offering limited access to discourse-level context and event structure
\cite{Huth2016,Jain2018}. Recent advances in large language models
provide a way to bridge this divide. By maintaining internal
representations that evolve with accumulated context and change
sharply at narrative boundaries, these models enable annotation-free
formalization of contextual drift and event shifts directly from
naturalistic input
\cite{Schrimpf2021,Caucheteux2022,Antonello2024,Michelmann2025}. Combined
with voxel-wise encoding and dense-sampling fMRI, this approach allows
a unified test of how distinct temporal regimes jointly support
coherence across cortical systems.

\begin{figure}[ht!]
  \centering \includegraphics[width=\linewidth]{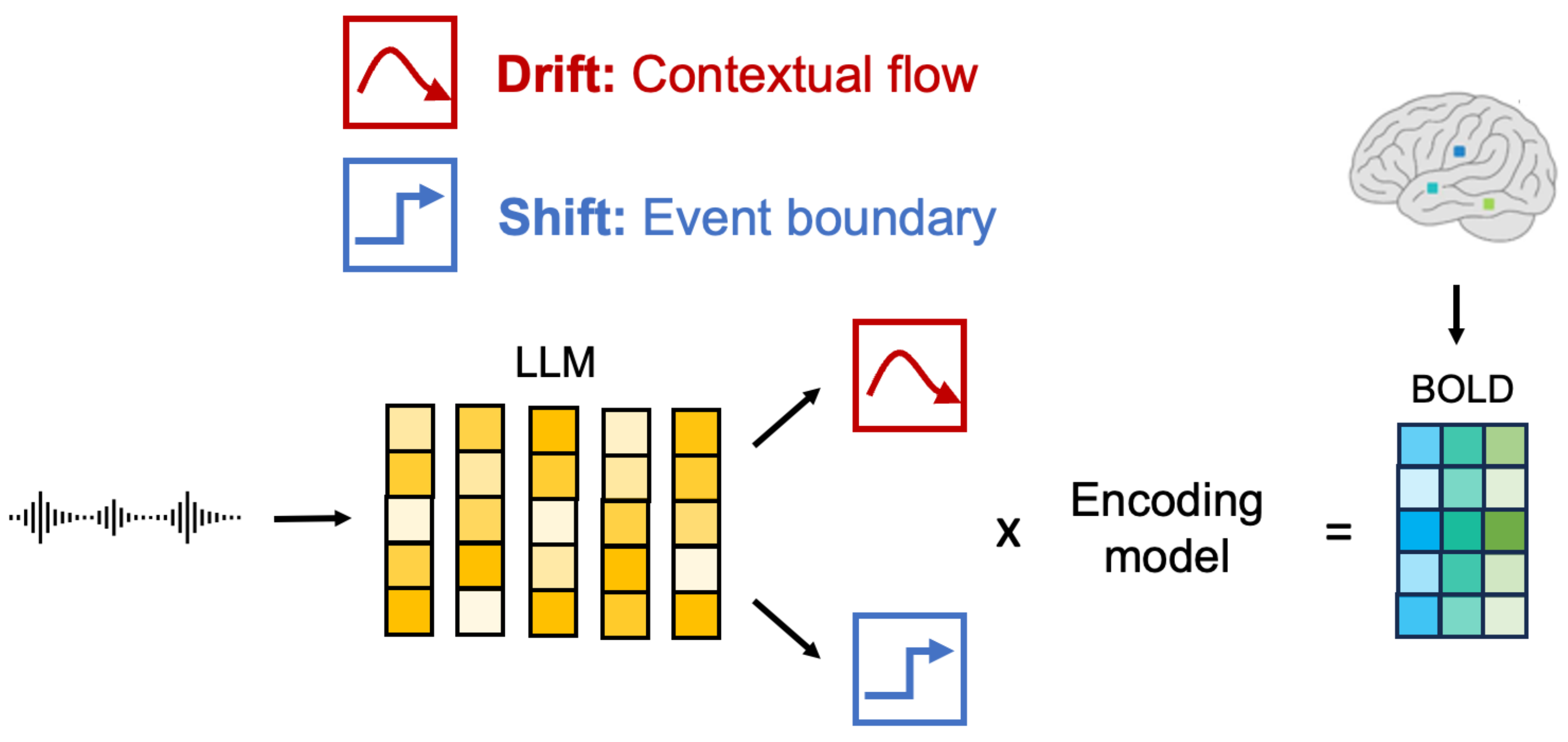}
\caption{\textbf{Annotation-free mapping of narrative coherence to
    brain dynamics.} Crime stories were processed by a decoder-only
  large language model (LLM) to derive two complementary signals of
  narrative structure: \emph{drift}, capturing the gradual accumulation of
  contextual meaning, and \emph{shift}, capturing discrete event
  boundaries. These LLM-derived signals were aligned to the spoken
  narratives presented during fMRI and entered into an encoding model
  to predict voxelwise BOLD responses. Comparing predicted and
  observed activity allowed us to map distinct neural systems
  associated with gradual contextual integration and rapid
  event-driven reconfiguration across the cortex.}
  \label{fig:llm_schema}
\end{figure}

Here, we combined model-derived drift and shift signals with
high-precision fMRI during extended narrative listening. We densely
sampled a single healthy participant across multiple hours of story listening,
enabling stable voxel-wise encoding models and out-of-sample
validation at the level of entire narratives
(Figure~\ref{fig:llm_schema}). We hypothesized that drift signals
would be expressed more strongly in association networks with long
integration windows, including default-mode and parietal regions,
whereas rapid shift signals would preferentially engage auditory and
language-selective regions sensitive to boundary-related
reorganization. By partitioning explanatory variance between these
processes within a single model-based framework, this work advances a
mechanistic account of coherence in language comprehension and
provides a principled foundation for probing coherence disturbances in
psychiatric conditions.

\section*{Materials and Methods}\label{sec:methods}

\subsection*{Sample size and scanning schedule}
One healthy, right-handed, male, native German speaker (age=28 years)
was densely sampled across eight scanning sessions spanning 43 days
(16~April--29~May~2024). Across all sessions we collected 16 narrative
runs; 3 runs were excluded a priori (1 due to excessive motion; 2 due
to signal loss), leaving 13 runs for analysis (one story per
run). Exact dates, start times, run counts, and story assignments are
listed in Table \ref{app:stories-schedule}.

\subsection*{Stimuli}\label{sec:stimuli}
The participant listened passively to 13 narrated,
single-voice German renditions of classic crime short stories (one per
run) from the 19th/early 20th century (median 31.5\,min;
range 13.8–55.0\,min). Full titles, authors, and per-scan
assignments are listed in Table \ref{app:stories-schedule}. Audio was
delivered via MR-compatible insert earphones (Optoacoustics FOMR
III+). Each run contained one continuous story (no task). We used
leave-one-story-out cross-validation at the level of stories (each
story serving once as the held-out test set; Supplementary
Methods). To reduce onset/offset transients we trimmed training
stories by 20 TRs at both onset and offset and test stories by 30 TRs
at onset and 20 TRs at offset.

\subsection*{Text-audio alignment (overview)}\label{sec:alignment}
We force-aligned the story text to audio at the word level using the
\texttt{torchaudio} MMS pipeline \cite{Pratap2023,PyTorchAudio2024} on
romanized/normalized text, producing word start/end times (Praat
format). Text-derived features were converted to continuous time and
resampled to the TR grid. Full alignment and resampling details
(including the resampling kernel) are provided in
\ref{app:expanded-methods-align}.

\subsection*{LLM-derived signals (overview)}\label{sec:features}
We derived two \emph{annotation-free} signals from the same
decoder-only LLM
(Llama-3.3-70B-Instruct\cite{Grattafiori2024}). \emph{Shift} is the
per-word log-probability that the next generated symbol is a special
boundary marker; \emph{drift} is the magnitude of change in a
leaky-integrated hidden-state trace. Both signals were aligned to word
times, resampled to TR, $z$-scored within story, and expanded into a
14-lag FIR design. The exact prompt,
layer/windowing choices, drift definition, and the boundary feature
bank are given in Supplementary Methods.

\subsection*{Data acquisition and preprocessing (overview)}\label{sec:preproc}
Data were acquired on a 7T Siemens MAGNETOM Terra with a 32-channel
head coil. Functional data used 2D gradient‑echo EPI (TR
$=1.18$\,s). Structural data were acquired using an MP2RAGE
sequence. Prior to preprocessing, MP2RAGE images underwent background
noise suppression. Preprocessing was performed using fMRIPrep 21.0.1
\cite{Esteban2019}, including motion correction and coregistration to
the T1w image; slice-timing correction was skipped. Post-processing
steps were implemented in custom Python scripts, comprising nuisance
regression of 22 confounds (6 motion parameters with derivatives,
framewise displacement, global signal with derivative, CSF, white
matter, and 5 anatomical CompCor components). High-pass filtering
($>0.0078$\,Hz) was implemented via discrete cosine transform basis
functions during regression, followed by low-pass filtering
($<0.1$\,Hz) using a Savitzky-Golay filter. Full acquisition and
preprocessing parameters are provided Supplementary Methods. Mean
frame-wise displacement per story/run is shown in
Figure~\ref{fig:app_md_displacement}.

\subsection*{Regions of interest}\label{sec:rois}
Regions of interest were anatomically defined from the probabilistic
Harvard–Oxford cortical and subcortical atlas \cite{Desikan2006},
split into left/right components and transformed to native functional
space for signal extraction.

\subsection*{ROI groups used in summary figures}\label{sec:roi-groups}
To summarize replicability across anatomically related regions, we
organized Harvard–Oxford ROIs into two \emph{a priori} sets: the
peri-Sylvian language network (LANG) and the
default-mode/parietal-integration network (DMN–PI), as specified
below. These sets were used only for aggregation and display.

\subsubsection*{Language network (peri-Sylvian; LANG)}
This set comprises bilateral peri-Sylvian speech/language cortex,
specifically the following Harvard–Oxford regions: Heschl’s gyrus
(including H1/H2), planum temporale, planum polare, superior temporal
gyrus (anterior and posterior divisions), middle temporal gyrus
(anterior, posterior, and temporo‑occipital divisions), inferior
frontal gyrus (pars opercularis and pars triangularis), the
frontal/central/parietal operculum, supramarginal gyrus (anterior and
posterior), insular cortex, and temporal pole.

\subsubsection*{Default-mode/Parietal-integration network (DMN–PI)}
This set comprises core default-mode hubs together with
dorsal/posterior parietal regions that support long-timescale
integration and integrative attention, namely: angular gyrus,
precuneus cortex, posterior cingulate (Harvard–Oxford "Cingulate
Gyrus, posterior division"), medial prefrontal/paracingulate cortex
(Harvard–Oxford "Frontal Medial Cortex" and "Paracingulate Gyrus"),
and superior parietal lobule. We use the label \emph{DMN–PI}-rather
than "DMN"-to make explicit that superior parietal lobule (a dorsal
attention/control region) is included by design.

\subsection*{Voxelwise encoding models (overview)}\label{sec:encoding}
For each voxel, we fit a ridge-regularized FIR encoding model mapping
lagged features to BOLD responses on concatenated training
stories. The held-out story remained untouched until
evaluation. Hemodynamic delays were modeled with 14 lags
(1.18–16.5\,s). Model selection used bootstrap cross-validation on
temporally contiguous chunks. Full design choices and hyperparameters
are detailed in Supplementary Methods.

\subsection*{Evaluation and statistical inference (overview)}\label{sec:stats}
Predictive evidence was quantified as the correlation between held-out
predictions and BOLD responses, with significance assessed at the
region-of-interest (ROI) level using a block-permutation null and
Simes omnibus test. For cross-story stability, we report the number of
stories in which the ROI-level Simes test reached significance
($p < .05$), displayed on the cortical surface and within key anatomical
ROIs. To map effects at the voxel level, we used cluster-based
correction. Specifically, we regressed held-out BOLD responses onto
the two predictions to isolate the unique contributions of shift and
drift. The resulting statistical maps of mean coefficients, along with
the directional contrast map
($\Delta\beta=\beta_{\text{shift}}-\beta_{\text{drift}}$), were
thresholded using a cluster-defining threshold of $Z > 2.3$ and a
cluster-extent significance level of $p < .05$. Voxelwise prevalence
within ROIs was additionally assessed by calculating the proportion of
voxels per ROI surviving the $Z > 2.3$ threshold. For detailed
procedures, see Supplementary Methods.

\subsection*{Text-derived punctuation control (overview)}\label{sec:punct_regressor}
Mechanistically, event boundaries in speech often coincide with
prosodic "edge" cues (pauses, pitch resets, durational changes) that
elicit boundary-locked responses in bilateral auditory cortices,
including Heschl's gyrus and STG
\cite{Ischebeck2008,Meyer2002,Steinhauer2001,Anurova2022}. Because
such prosodic edges often co-occur with punctuation, we also included
a control analysis that tests whether an LLM-based, shift-like signal
is reducible to punctuation/prosody. We constructed a punctuation
regressor from the marks \texttt{.,:;} placed at the offsets of the
corresponding words and evaluated a pairwise stage‑2 model (unique
\emph{shift} vs.\ unique punctuation) on held-out stories. The
construction and estimation details are provided in Supplementary
Methods.

\subsection*{Timescale sweep for drift (overview)}\label{sec:rhoshort}
To understand how long past information influences current processing,
we systematically varied the integration timescale using a `leak'
parameter ($\rho$). This parameter controls the rate of decay: a low
$\rho$ means that past context fades quickly, while a high $\rho$
allows information to persist over longer periods. We tested a wide
range of values ($\rho \in {0.01, \dots, 0.90}$), re-calculating the
drift signal and repeating the evaluation for each setting. We then
summarized the results across regions by tracking how consistently the
model predicted brain activity across different stories.

\section*{Results}\label{sec:results}

\begin{figure}[h!]
  \centering
  \includegraphics[width=\linewidth]{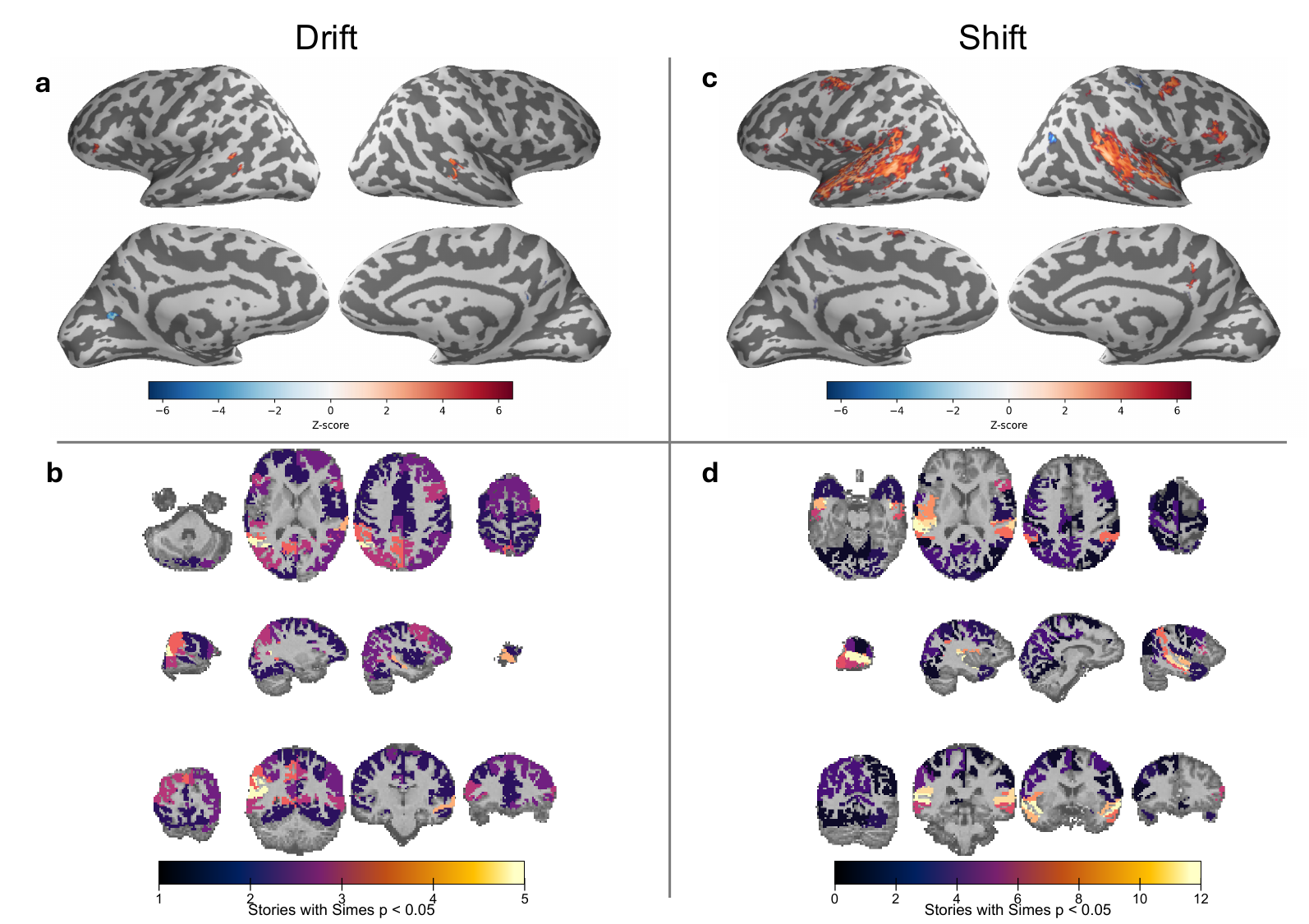}
\caption{\textbf{Across stories activation and consistency
    maps for drift and shift.}  \textbf{a)} Cortical surface
  projections showing Z-scores for drift.  \textbf{b)} Volumetric
  region of interest maps showing the cross-story consistency of
  effects, where color intensity represents the number of stories (out
  of 13) in which each region was significant (Simes $P < 0.05$). The
  drift maps showed comparatively weaker and more distributed patterns
  with lower consistency, involving higher-level hubs like the angular
  gyrus and precuneus (default mode network), consistent with
  integrative or attentional components rather than primary auditory
  drive. \textbf{c)} Cortical surface projections displaying Z-scores
  for shift. \textbf{d)} The shift maps exhibited robust, bilateral
  modulation within classic speech areas around the Sylvian fissure
  (e.g., Heschl’s gyrus and STG), showing high consistency across up
  to 13 stories.}
  \label{fig:consistency_maps}
\end{figure}

\subsection*{Dissociable cortical systems encode drift and shift signals}
Drift and shift signals exhibited dissociable whole-brain predictive
topographies, with drift effects distributed across the heteromodal
association cortex, and shift effects concentrated in peri-Sylvian
language regions (Figure~\ref{fig:consistency_maps}). To evaluate how
well these effects generalized across narratives, we evaluated
cross-story consistency by region of interest. For this, we used
region of interest-level \emph{Simes counts}, defined as the number of
stories (out of 13) in which a region showed a significant effect
(\hyperref[sec:methods]{Materials and
  Methods}). Figure~\ref{fig:consistency_maps}b summarizes these counts
across the Harvard--Oxford atlas \cite{Desikan2006}. More
specifically, for \emph{drift}, effects showed that replicability
concentrated in heteromodal association hubs. The most consistent
effects were observed in angular gyrus, precuneus, posterior
cingulate, and superior parietal regions, whereas early auditory
cortex rarely showed replicable drift effects under identical
criteria. Thus, drift generalized preferentially in default-mode and
parietal hubs rather than in auditory core regions
(Figure~\ref{fig:consistency_maps};
Table~\ref{tab:roi_simes_combined}). By contrast, \emph{shift} effects showed a
complementary profile. They generalized robustly across narratives
within peri-Sylvian speech regions. Core auditory–language regions
reached ceiling or near-ceiling consistency, including Heschl’s gyrus,
planum temporale/polare, and superior temporal cortex
bilaterally. These patterns indicate highly reproducible,
boundary-linked responses to shift across stories
(Figure~\ref{fig:consistency_maps}).

\subsection*{Distinct predictive contributions of drift and shift}
\begin{figure*}[h!]
  \centering
  \includegraphics[width=\linewidth]{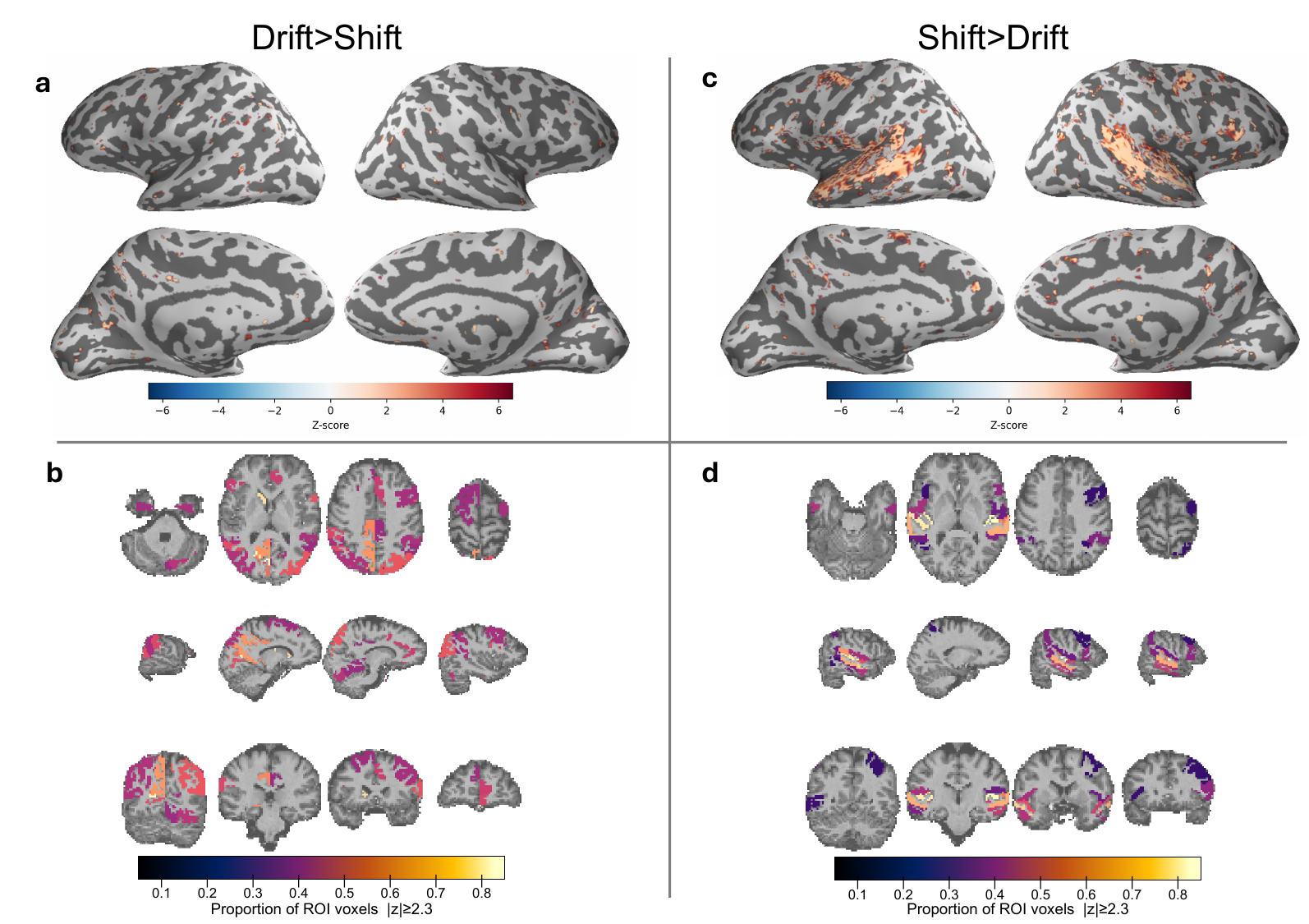}
  \caption{\textbf{Unique predictive contributions of drift and
      shift.}  \textbf{a)} Voxelwise significance map for unique drift
    effects, localized primarily to heteromodal association hubs,
    including angular gyrus and precuneus.  \textbf{b)} Corresponding
    ROI-level effect sizes for drift in posterior-inferior
    default-mode regions (DMN–PI).  \textbf{c)} Voxelwise significance
    map for unique shift effects, estimated after regressing out
    shared variance with drift. Effects concentrate bilaterally in the
    peri-Sylvian language network, including Heschl’s gyrus, superior
    temporal cortex, and planum temporale/polare.  \textbf{d)}
    ROI-level effect sizes for shift in language-network ROIs (LANG),
    showing mean unique regression weights ($\beta$) across 13 stories
    with 95\% bootstrap confidence intervals.}
  \label{fig:delta_beta}
\end{figure*}

Because drift and shift are correlated, we isolated their independent 
contributions by fitting a joint model and testing the partial voxelwise 
coefficients across held-out stories (Materials and
Methods). Results are summarized by voxelwise significance maps,
region-level effect sizes, and the directional contrast
$\Delta\beta=\beta_{\text{shift}}-\beta_{\text{drift}}$
(Figure~\ref{fig:delta_beta}).

\subsubsection*{Drift predictions are dominant in heteromodal association hubs}
Controlling for shift, drift independently predicted activity in higher-order association regions,
most consistently in angular gyrus, precuneus, posterior cingulate,
and medial prefrontal cortex (Figure~\ref{fig:delta_beta}). All these 
regions can be broadly summarized as Default-Mode Network/Parietal Integration (DMN-PI).
In these hubs, the directional contrast $\Delta\beta$ was small, consistent
with drift contributing reliably but not overwhelmingly relative to
shift.

\subsubsection*{Boundary-linked shifts dominate in peri-Sylvian language responses}
Unique shift effects concentrated bilaterally in the peri-Sylvian
language belt, including Heschl’s gyrus, planum temporale and polare,
anterior and posterior superior temporal cortex, posterior
supramarginal gyrus, and inferior frontal gyrus (pars triangularis)
(Figure~\ref{fig:delta_beta}). These regions can be broadly clustered 
into a Language network (LANG). Accordingly, $\Delta\beta$ was positive
throughout theis network, indicating that boundary-linked shifts yield 
stronger predictions than drift within canonical speech-processing 
cortex, even when both processes co-occur.

\subsubsection*{Secondary drift-dominant effects outside the language system}
Outside the language network, several sensory–motor and visual regions
exhibited negative $\Delta\beta$ values (drift $>$ shift), including
somatosensory cortex and occipital areas, indicating stronger unique
contributions of slowly varying contextual signals in these processing
streams (Table~\ref{tab:roi_unique_all}).

\subsubsection*{Effect sizes and coefficient sign}
To quantify magnitude beyond significance, we report mean unique
regression weights across stories with 95\% bootstrap confidence
intervals for representative ROI sets (DMN–PI for drift, LANG for
shift; Figure~\ref{fig:delta_beta}). Because predictors and BOLD
signals were z-scored within story and modeled with Finite Impulse Response (FIR) lags,
coefficient sign reflects relative phase alignment rather than
excitation versus inhibition; interpretation therefore focuses on
spatial distribution and relative magnitude of unique effects.

\subsection*{Drift depends on a finite integration timescale}\label{sec:rho}
To characterize the temporal window over which contextual information
contributes to drift, we varied the memory decay parameter $\rho$ and
recomputed drift predictors across a wide range of integration
timescales. For each value of $\rho$, we evaluated cross-story
replicability using the same voxel-wise inference and ROI-level Simes
aggregation as in the main analyses (Materials and Methods).

\begin{figure}[H]
  \centering
  \includegraphics[width=\linewidth]{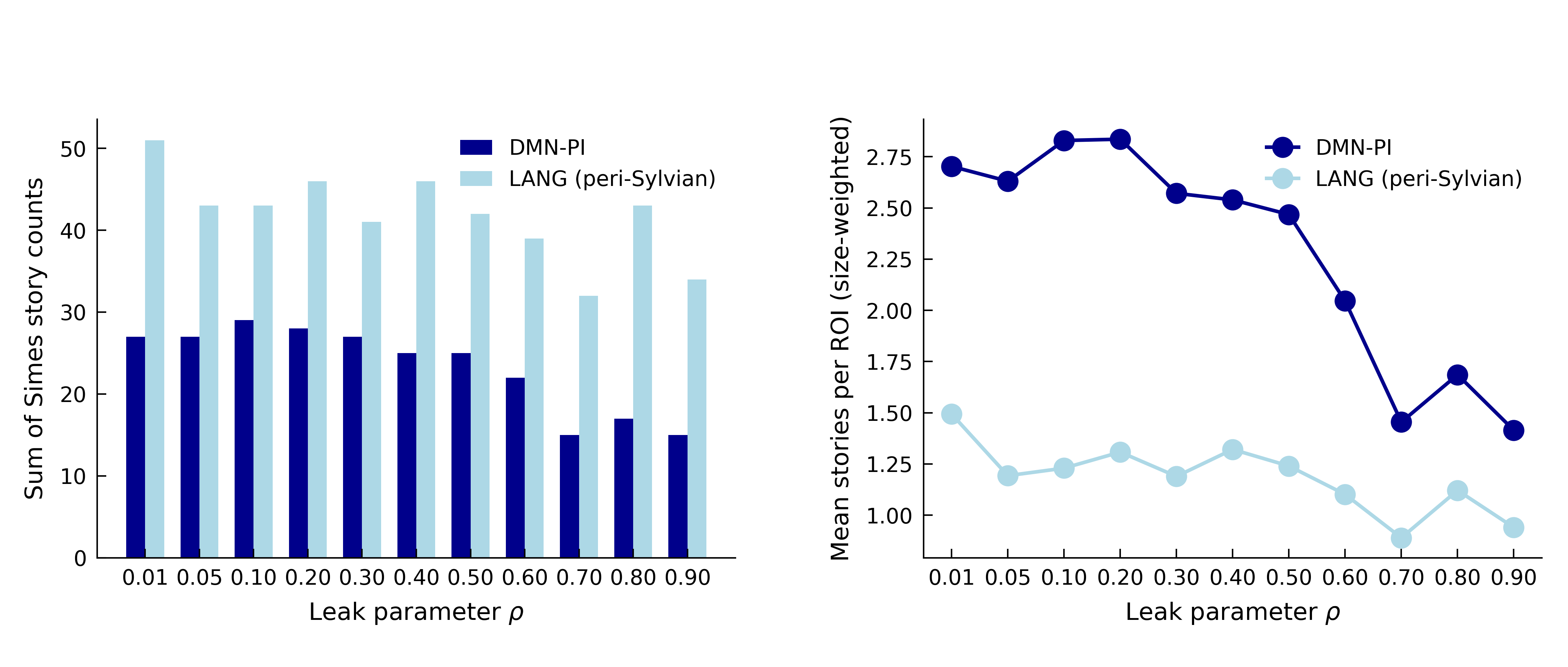}
\caption{\textbf{Replicable drift effects across integration
    timescales.}  Cross-story generalization of drift effects as a
  function of the memory decay parameter $\rho$ \textbf{a)} Raw sums
  of region of interest-level Simes significant-story counts (not
  normalized for ROI number or size). \textbf{b)} Size-weighted mean
  significant stories per region, normalizing for both region number
  and size. \emph{After normalization}, drift generalized more strongly in
  posterior-inferior default-mode and parietal regions (DMN–PI) than
  in the language network (LANG), with a broad maximum at intermediate
  integration timescales ($\rho\approx0.10$–$0.20$).}
  \label{fig:rho_simes}
\end{figure}

Across timescales, drift effects generalized more strongly in
posterior-inferior default-mode and parietal regions (DMN–PI) than in
the language network (LANG)
(Figure~\ref{fig:rho_simes}). Replicability peaked at intermediate
integration timescales ($\rho\approx0.10$–$0.20$), whereas very short
timescales yielded near-memoryless traces with weak generalization and
very long timescales over-smoothed the signal. Thus, drift reflects
integration over a finite temporal window, rather than unlimited
accumulation, with a preferred timescale in the heteromodal
association cortex.

\subsection*{Punctuation control: shift predicts boundary responses beyond pauses}\label{sec:results_punct}
To test whether the shift signal merely reflects low-level pauses or
typography, we included punctuation (\texttt{.,:;}) as a competing
predictor. When entered into the joint model, shift retained large and
focal unique effects across bilateral peri-Sylvian regions, including
Heschl’s gyrus, planum temporale/polare, and superior temporal cortex,
whereas punctuation showed markedly smaller unique effects in the same
areas (Figure~\ref{fig:app_punct_unique_zmaps}). Moreover, punctuation
exhibited limited cross-story generalization, reaching significance in
only a small subset of ROIs and stories, in contrast to the robust and
widespread replicability observed for shift. These results indicate
that shift captures boundary-linked neural responses that cannot be
explained by pauses or typographic markers alone.

\subsection*{Summary}
Together, these analyses showed that narrative comprehension is driven 
by an interaction between a slow context-integrating signal in 
heteromodal association cortex and a fast boundary-linked signal 
in peri-Sylvian regions.

\section*{Discussion}
In this study, we demonstrated that narrative coherence arises from 
the dynamic interplay of slow semantic evolution (drift) and rapid 
event transition (shift), mapping onto distinct yet overlapping 
cortical systems. Both signals were derived annotation-free
from a single large language model and predicted BOLD responses during
extended story listening, yet they dissociated systematically across
cortical networks. Drift generalized preferentially in heteromodal
association hubs, including default-mode and parietal cortex, whereas
boundary sensitivity was dominant in peri-Sylvian auditory–language
regions. Crucially, unique-effects analyses revealed that these
processes are not mutually exclusive: higher-order hubs such as
angular gyrus and precuneus expressed sensitivity to both signals,
suggesting that coherence emerges from their coordinated interaction
rather than from a single mechanism.

These findings situate coherence at the intersection of two lines of
work that have often been treated separately: gradual context
integration and event segmentation
\cite{Zacks2007,Kurby2008,DuBrow2017,Baldassano2017,Chen2017}. By
formalizing both operations within a single, stimulus-derived
framework, our results reconcile perspectives that emphasize discrete
event boundaries with those that stress continuous representational
drift. The observation that default-mode hubs track both drift and
shift aligns with proposals that these regions maintain situation
models across time while remaining sensitive to structural changes
that delimit events.

Our results also map naturally onto hierarchical temporal receptive
window (TRW) accounts of naturalistic comprehension, which posit
progressively longer integration windows from early auditory cortex to
association cortex
\cite{Hasson2008a,Lerner2011,Honey2012,Simony2016}. Within this
framework, rapid boundary-linked responses are expected to peak in
peri-Sylvian speech regions, whereas slower accumulation of contextual
meaning should preferentially engage posterior and medial association
hubs. Consistent with this view, drift effects generalized most
strongly in posterior-inferior default-mode and parietal regions and
peaked at intermediate integration timescales, indicating that these
regions integrate context over a finite window rather than
accumulating information indefinitely. This timescale tuning provides
a concrete instantiation of TRW hierarchies using explicit,
model-based predictors \cite{Chang2022,Geerligs2022}.

Recent work has shown that transformer-based language models robustly
predict neural responses during language comprehension and align
hierarchically with cortical processing stages
\cite{Schrimpf2021,Caucheteux2022,Goldstein2022}. Importantly, such
alignment does not imply that next-token prediction alone captures
human comprehension \cite{Antonello2024}. Our results extend this
literature by showing that LLM-derived internal dynamics can be used
to formalize distinct temporal components of coherence itself. In
particular, recent studies demonstrate that large language models can
segment narrative events in a manner comparable to human judgments
\cite{Michelmann2025}, providing an external validation for the use of
LLM-based shift signals as proxies for narrative structure.

Data-driven event segmentation approaches applied directly to neural
activity, such as hidden Markov models, have reported hippocampal
responses time-locked to inferred event boundaries
\cite{Baldassano2017,Chen2017,BenYakov2018}. We did not observe 
reliable hippocampal effects in the present data. This absence 
may reflect the temporal sparsity of hippocampal boundary responses 
and their sensitivity to memory encoding and retrieval demands. 
Critically, our study was not explicitly optimized or adequately 
powered for hippocampal effects, given the dense-sampling rationale 
prioritized cortical networks, and subcortical measurement at 
ultra-high field is often hindered by signal loss and 
susceptibility artifacts. Consequently, targeted acquisition 
designs tailored to subcortical coverage will be needed to test
hippocampal contributions to narrative boundaries more directly. \cite{Silva2019}.

Event boundaries in speech are often accompanied by prosodic cues such
as pauses, pitch resets, and durational changes, which elicit robust
responses in early auditory cortex
\cite{Steinhauer2001,Meyer2002,Ischebeck2008}. Neurophysiological and
neuroimaging studies show that such acoustic edges drive strong
boundary-locked responses in Heschl’s gyrus and superior temporal
cortex, even in the absence of lexical content
\cite{Giraud2012,Ding2012,Forseth2020a,Anurova2022}. Our control
analyses showed that a punctuation-based regressor captured a limited
portion of this variance, but that the LLM-derived shift signal
explained substantial, focal variance beyond punctuation alone. This
dissociation indicates that shift captures higher-order boundary
structure rather than merely reflecting pauses or typographic markers.

Several limitations warrant consideration. First, the dense-sampling design involved a single healthy participant. This was an intentional methodological choice: single-subject depth reveals intra-individual effect sizes that are systematically obscured in group averages, and precision phenotyping of this kind provides a powerful platform for hypothesis generation prior to population-level scaling \cite{DeYoung2025,Tahedl2026}. That said, whether these individual-level effects replicate across participants and populations remains an open question, and multi-participant studies are the explicit next step of this research program. Second, drift and shift signals were derived from a single decoder-only LLM (Llama-3.3-70B-Instruct). Robustness across model families, architectural scales, and languages has not yet been systematically assessed. However, the anatomical specificity of the effects and their consistency across 13 independent narratives argue against the findings being attributable to model-specific idiosyncrasies or training data artifacts; replication with alternative architectures will nonetheless be important. Third, the passive listening paradigm did not include concurrent behavioral comprehension measures, which means we cannot directly dissociate active narrative tracking from passive auditory processing. Future work should incorporate online comprehension probes to establish the behavioral correlates of the encoding effects reported here. Finally, acoustic and prosodic features were not explicitly modeled, and residual variance in early auditory regions, particularly around Heschl's gyrus, may partly reflect such low-level signals. The punctuation control analysis, which showed that the shift signal retains large focal effects beyond typographic boundary markers, partially addresses this concern, but dedicated acoustic modeling remains an important extension.

While our results address basic systems neuroscience, the dissociation between gradual 
context integration (drift) and rapid event-driven reconfiguration
(shift) provides a powerful translational framework. Formal 
thought disorder, a core feature of psychosis, manifests 
as a profound disturbance in discourse organization \cite{Cavelti2018a,Winkelbeiner2018} often presenting with either excessive semantic wandering (derailment) 
or abrupt loss of context (tangentiality). By demonstrating 
that coherence relies on these two temporally distinct, model-derived 
signals, our findings offer a mechanistic entry point to assess 
such clinical phenomena. Specifically, this framework hypothesizes 
that derailment may reflect an impairment in context-integrating 
drift signals within the default-mode network, whereas abrupt topic 
shifts might stem from a failure in boundary-linked shift detection 
in auditory-language networks. Future translational work should 
apply this model-based feature space to clinical populations to 
probe precisely which temporal regime is selectively disrupted in 
psychosis, outlining concrete translational next steps for 
biomarker discovery.

\bibliographystyle{elsarticle-num}
\bibliography{Staub2025}

\clearpage 

\section*{Author contributions statement}
F.R., N.L. and P.H. conceived the experiment,  F.R. and N.L. conducted the 
experiment, D.S., F.R., A.M., Y.P., R.H., N.L. analyzed the results.
D.S., F.R., A.M., Y.P., and P.H. conceived the manuscript. All authors reviewed the manuscript. 

\section*{Declaration of competing interest}
PH has received grants and honoraria from Novartis, Lundbeck, Mepha,
Janssen, Boehringer Ingelheim, Neurolite, and OM Pharma outside of
this work. No other disclosures were reported.

\section*{Ethics approval and consent to participate}
The study was approved by the local ethics committee (KEK-ZH
2024-01314). All procedures complied with the Declaration of
Helsinki. Written informed consent was obtained from the participant.

\section*{Data and code availability}
All analysis code and derived regressors will be made available at
\url{https://github.com/homanlab/brainencode/}.

\clearpage

\section*{Supplementary Information}

\setcounter{page}{1}
\setcounter{table}{0}
\setcounter{figure}{0}
\setcounter{equation}{0}

\renewcommand{\thepage}{S\arabic{page}}
\renewcommand{\thefigure}{S\arabic{figure}}
\renewcommand{\thetable}{S\arabic{table}}
\renewcommand{\theequation}{S\arabic{equation}}


\section*{Supplementary Methods}
\label{app:expanded-methods}

\subsection*{MRI acquisition and preprocessing (full)}\label{app:expanded-methods-acq}
Scanner: 7T Siemens MAGNETOM Terra; 32-channel head coil. Functional
scans: 2D gradient‑echo EPI, TR $=1.18$\,s; TE $=25$\,ms; flip angle
$45^\circ$; voxel size $2.3\times 2.3\times 2.3$\,mm$^3$ with 0.6\,mm
slice gap; multiband factor 3. Anatomy: T1-weighted multi-echo MPRAGE
(TR $=6$\,s; TE $=1.99$\,ms; voxel size 0.68\,mm
isotropic). Fieldmaps: 2D gradient-echo (TR $=2.25$\,s; TE $=3.06$\,ms
and $5.52$\,ms; flip angle $25^\circ$; matrix $64\times 64$; slice
thickness 4\,mm).

For preprocessing, fMRIPrep \cite{Esteban2019} was used; standardized
motion/distortion/slice-timing corrections and cross-run
coregistration using FLIRT \cite{Jenkinson2002} to the first run of
the session as reference were employed. Nuisance regression removed
variance associated with six motion parameters, their first
derivatives, and framewise displacement. Time series were low-pass
filtered (fifth-order, zero-phase Butterworth, 0.1\,Hz). Run-specific
brain masks were applied, and data were voxelwise $z$-scored within
run.

\subsection*{Text-audio alignment and resampling (full)}\label{app:expanded-methods-align}
For forced alignment, we romanized/normalized the German text
(lowercase, remove diacritics/punctuation) and applied the
\texttt{torchaudio} MMS aligner \cite{Pratap2023,Hwang2023} to obtain
Praat word start/end times. For each story, a \emph{DataSequence}
stored the word sequence, word timestamps, and the run-specific TR
grid; timestamps were shifted by the measured audio onset. Resampling
to TR was obtained by forming continuous-time feature traces by
convolving impulse trains at word times with a Lanczos (sinc-windowed)
interpolation kernel \cite{Lanczos1956}.

\[
L_a(x)=
\begin{cases}
\sinc(x)\,\sinc\!\big(x/a\big), & |x|<a,\\
0, & \text{otherwise},
\end{cases}
\]
with window parameter $a{=}1$ for shift and $a{=}3$ for drift, and sampled at TR centers.

\subsection*{LLM features and signals (full)}\label{app:expanded-methods-llm}
Model and readout: Decoder-only LLM:
\emph{Llama-3.3-70B-Instruct} (HuggingFace), greedy decoding, 8-bit
loading via \texttt{BitsAndBytes} \cite{Dettmers2022}. Hidden states
were read from layer $L{=}79$ using sliding token windows. For
boundary log-probabilities, the look-back window was $K{=}512$
tokens. For base features (hidden states) we used a one-word stride
with a variable look-back that ramps from 256 to 512 tokens and
resets, amortizing compute (an engineering choice; longer discourse
dependencies are governed by the drift integration
below).

Drift: Define hidden state $\mathbf{H}_t$, leaky integration
$\mathbf{g}_t=\rho\,\mathbf{g}_{t-1}+(1-\rho)\,\mathbf{H}_t$ with
$\mathbf{g}_0=\mathbf{H}_0$ and $\rho\in(0,1)$; drift magnitude
$\delta_t=\lVert \mathbf{g}_t-\mathbf{g}_{t-1}\rVert_2$. The per-TR
drift series was $z$-scored within story after trimming and before FIR
expansion. Unless otherwise stated, $\rho{=}0.30$.

Shift (boundary likelihood): We prompted the LLM to copy the text and
insert a single special marker "\verb|¶|" only at event boundaries
\cite{Michelmann2025}. At each word~$i$ we computed $\ell_i=\log
p(\text{¶}\mid\text{context}_{\le i})$. Because boundary scoring used
punctuated text while alignment used normalized text, we reconciled
positions with a sequence matcher and linearly interpolated unmatched
gaps.

Boundary feature bank: From the per-TR series $\ell(t)$ we built four
channels: (i) raw log-probability $x_1(t)=\ell(t)$; (ii) half-wave
derivative $x_2(t)=\max\{\ell(t)-\ell(t-1),0\}$; (iii) moving-average
high‑pass $x_3(t)=\ell(t)-(\ell*k_W)(t)$ with rectangular kernel $k_W$
of width $W{=}12$ TRs; (iv) smoothed thresholded
$x_4(t)=(\mathcal{G}_\sigma * r)(t)$ with $\sigma{=}1$ TR and
$r(t)=\max\{\ell(t)-\tau,0\}$, $\tau$ the 90th percentile of
$\ell(t)$. Each channel was $z$‑scored within story after trimming.

FIR expansion: For both drift (scalar) and shift (4-vector), we
stacked $D{=}14$ lags (1.18–16.5\,s), omitting lag~0.

\subsection*{Encoding model (full)}\label{app:expanded-methods-encoding}
Design and lags: FIR with $D{=}14$ lags (covering the
canonical positive BOLD lobe and early return to baseline). Lag~0 was
omitted to enforce causality and guard against residual alignment
error. \\

Estimation: Per-voxel ridge:
\[
\hat{\mathbf{w}}=\arg\min_{\mathbf{w}}\lVert \mathbf{y}-\mathbf{X}\mathbf{w}\rVert_2^2+\alpha\lVert \mathbf{w}\rVert_2^2,
\]
with $\alpha\in\{0\}\cup\{10^{-6},\ldots,10^{0}\}$ (49 log‑spaced
values). Model selection used bootstrap cross‑validation preserving
temporal contiguity (chunk length 40 TRs; typical validation fraction
$\approx$25\%). Alphas were selected by maximizing validation
correlation. The training design concatenated all training stories
after trimming; validation blocks could span story boundaries.

\subsection*{Evaluation and inference (full)}\label{app:expanded-methods-inference}

\subsubsection*{Held-out predictive evidence (marginal)}
For each voxel $v$ and test story $s$, we computed Pearson’s
correlation $r_{v,s}$ between the held-out prediction and the measured
BOLD time series. To obtain a null that preserves temporal
autocorrelation, we randomly permuted the order of contiguous 10‑TR
chunks of the test-story BOLD response (block permutation; $B{=}500$
shuffles), recomputing the correlation each time. The empirical
two‑sided voxelwise $p$-value was
\[
p_{v,s} \;=\; \frac{1 + \sum_{b=1}^{B} \mathbf{1}\!\big(\,|r_{v,s}^{(b)}| \ge |r_{v,s}|\,\big)}{B+1}.
\]
Within each ROI $R$, voxelwise $p$‑values $\{p_{v,s} : v \in R\}$ were combined using Simes’ omnibus test to yield an ROI-level $p_{\text{Simes}}(R,s)$.

\subsubsection*{Cross-story stability (ROI Simes counts)}\label{app:roi-simes-counts}
To quantify replicability across stories, we computed, for each ROI
$R$, the \emph{Simes count}
\[
C(R) \;=\; \sum_{s=1}^{13} \mathbf{1}\!\big(p_{\text{Simes}}(R,s) < 0.05\big).
\]
These counts summarize in how many held‑out stories an ROI showed
significant predictive evidence. They are descriptive of
\emph{replicable goodness-of-fit} (presence of an effect) and do not
encode effect sign or voxelwise spatial extent. No additional
voxelwise FDR is applied to the counts. Under the global null,
$\mathbb{E}[C(R)] = 13 \times 0.05 = 0.65$.

\subsubsection*{Unique-effects estimation (stage-2) and directional contrast}
To isolate independent contributions of the two predictors, we
regressed the measured BOLD onto the two held-out predictions for each
voxel $v$ and story $s$:
\begin{equation}
y_{v,s}(t)
= \beta_{v,s,\text{shift}}\,\hat{y}_{v,s}^{\text{shift}}(t)
+ \beta_{v,s,\text{drift}}\,\hat{y}_{v,s}^{\text{drift}}(t)
+ \varepsilon_{v,s}(t)
\label{eq:twostage}
\end{equation}

All time series were $z$-scored \emph{within story}; the intercept was
omitted (expected $\approx 0$). Coefficients were estimated by
ordinary least squares. For each voxel, we then tested across stories
whether the mean coefficient differed from zero (two‑sided one‑sample
$t$‑test, df$=12$) separately for $\beta_{\text{shift}}$ and
$\beta_{\text{drift}}$. Voxelwise $p$‑values were converted to $z$ for
display and controlled with BH–FDR at $q{=}0.05$ across cortical
voxels. ROI-level inference for unique effects used Simes on the
voxelwise $p$-values within each ROI.

We also formed the directional contrast
$\Delta\beta_{v,s}=\beta_{v,s,\text{shift}}-\beta_{v,s,\text{drift}}$
and analyzed it analogously (across-story one-sample $t$, voxelwise
BH–FDR $q{=}0.05$, plus ROI-level Simes).

\subsubsection*{Constants and implementation notes}
Block-permutation chunk length for the test story was 10 TRs; the
number of permutations was $B{=}500$. All tests were two‑sided. BH–FDR
control at $q{=}0.05$ was applied independently for each voxelwise map
(\(\beta_{\text{shift}}\), \(\beta_{\text{drift}}\), and
$\Delta\beta$). Simes tests were applied within ROIs to the relevant
voxelwise $p$-values. No additional cluster-extent threshold was
imposed on the voxelwise maps; small but FDR-significant clusters are
therefore expected. The Simes tests provide ROI-level inference by
combining voxelwise $p$-values within each anatomical ROI and do not
implement a separate cluster-size correction.

\subsection*{Controls and parameter sweep (full)}\label{app:expanded-methods-controls}
Punctuation control: From the punctuated text we flagged
tokens ending in \texttt{.,:;}. Flags were time-stamped at the
\emph{offset} of the preceding aligned word, converted to a per‑TR
series, $z$-scored within story, and expanded into a 14-lag
FIR. Stage-1 ridge produced held-out predictions. Stage-2
unique-effects estimation used the pairwise model
\[
y(t)=\beta_{\text{shift}}\hat{y}_{\text{shift}}(t)+\beta_{\text{punct}}\hat{y}_{\text{punct}}(t)+\varepsilon(t),
\]
tested across stories as in above.

Timescale sweep: For
$\rho\in\{0.01,0.05,0.10,\ldots,0.90\}$ we recomputed $\delta_t$,
rebuilt the design, and repeated evaluation. For each ROI we recorded
the number of test stories with Simes $p{<}.05$. We summarize
(\textit{i}) raw sums across ROIs and (\textit{ii}) size-weighted mean
significant stories per ROI (voxel-weighted) within LANG and DMN–PI.

\subsection*{Validating the LLM event-boundary metric}

\subsubsection*{Rationale}
Our boundary regressor is the LLM’s \emph{log-probability} of
inserting a special boundary token at each word position. If this
quantity truly captures event structure, then words labeled as event
boundaries by humans should, on average, receive \emph{higher} scores
than non-boundaries, and a threshold on the score should recover human
boundaries with few false alarms. We therefore evaluated the regressor
with complementary, threshold-free and discrete criteria.

\subsubsection*{Data and alignment}
All validation analyses in this appendix were conducted on a
\emph{separate} German crime story that we generated with GPT-4. We
prompted the model to create a narrative with \emph{clear,
well-delimited events} to obtain unambiguous candidate
boundaries. Nine fluent German raters (R1–R9) then read the text and
independently inserted an event marker immediately \emph{after} the
last word of any event they perceived as ending. Raters were blinded
to the LLM boundary scores and to each other’s markings. This stimulus
was \emph{not} used for any fMRI modeling or evaluation and serves
purely to test whether the LLM-derived boundary metric behaves as
intended on a controlled narrative. We aligned scores to a word-time
axis using a nominal reading rate (150 wpm), no audio or fMRI data
were used for this appendix analysis, and computed the boundary score
$\ell_i=\log p(\text{¶}\mid\text{context}_{\le i})$ at every
word. Unless stated otherwise, analyses in this appendix use tolerance
\(=\) \(\pm 0\) words around human boundaries, block bootstrap with
\SI{50}{word} blocks for CIs, and $n_{\text{boot}}=10{,}000$
resamples.

\subsection*{Evaluation strategy}
\subsubsection*{(1) Threshold-free discrimination}
We treated the per-word score as a continuous decision variable and
computed AUROC and AUPRC. AUROC reports the probability that a random
human boundary receives a higher score than a random non-boundary;
AUPRC reports precision vs.\ recall under class imbalance. We also
report $d'$ derived from AUROC via the equal-variance SDT link \(
d'=\sqrt{2}\,\Phi^{-1}(\text{AUROC}) \) and, secondarily, a parametric
$d'$ from the positive/negative score distributions.

\subsubsection*{(2) Consensus sensitivity}
To relate model scores to annotation strength, we recomputed AUROC
while requiring that positives be marked by at least \(k\) raters (
$\geq$1, $\geq$2, ...). We show 95\% CIs via block bootstrap and the
count of positives at each threshold.

\subsection*{Results for the synthetic validation story}
For our synthetic story (1{,}256 words; positives 49; base rate
0.039), we obtained:
\begin{itemize}
  \item \textbf{AUROC} = \textbf{0.9448} \; (95\% CI
    \textbf{0.8991–0.9751}; \SI{50}{word} blocks,
    $n_{\text{boot}}=10{,}000$), corresponding to \(\boldsymbol{d'}\)
    (from AUC) \(=\) \textbf{2.257}.
  \item \textbf{AUPRC} = \textbf{0.4283} versus a class-balance baseline of \textbf{0.0390} (\(\sim\)11× lift).
  \item \textbf{Distributional} \(d'\) (equal-variance SDT) =
    \textbf{2.849}.
  \item \textbf{Consensus link}: Spearman \(\rho\) between score and
    number of raters = \textbf{0.299}; AUROC remains \(\gtrsim 0.95\)
    for mid-to-high consensus thresholds (counts per threshold are
    shown below).
\end{itemize}
These converging results demonstrate that the regressor is highly
informative about human boundaries without manual
thresholds. Together, they verify that the boundary metric behaves
exactly like it is supposed to.

\subsection*{Control for punctuation}
A potential confound is that event boundaries often coincide with
sentence punctuation; if so, our boundary regressor might merely
reflect punctuation. We therefore quantified (i) the association
between punctuation and the LLM score, (ii) how well punctuation alone
predicts human event boundaries, and (iii) whether the LLM score
remains predictive \emph{beyond} punctuation,both \emph{within
punctuation strata} and after removing the punctuation
\emph{mean-level shift}.

\subsubsection*{Operationalization}
We flagged a word as “punctuation” if, after stripping closers
(quotes/brackets), it ended in one of \{.\,!\,?\,…\,;\,:\,,\}. Using
the same word grid as the validation, we compared this indicator to
the LLM boundary score $\ell_i=\log p(\text{¶}\mid\text{context}_{\le
  i})$ and to the human event-boundary labels (union across raters;
tolerance $=\pm 0$ words).

\subsubsection*{Mean-level shift}
We observed that punctuation words have a higher \emph{average} LLM
score than non-punctuation words (here, means $-7.29$
vs.\ $-12.93$). This global offset is the \emph{mean-level shift}. To
test whether discrimination is driven only by this offset, we
\emph{residualized} the score by centering within groups:
$s_{\mathrm{resid}} = s - \mu_g$ with
$g\in\{\text{non-punct},\text{punct}\}$ and $\mu_g$ the mean score in
group $g$. Residualizing removes the global two-mean difference; any
remaining separation reflects \emph{within-group} information.

\subsubsection*{Within punctuation strata}
We also assessed discrimination \emph{within} each punctuation subset,
i.e., among \emph{only} non-punctuation words and among \emph{only}
punctuation words. This conditions on punctuation status so that
punctuation itself cannot explain any separation. If the LLM score
still distinguishes human EBs in either subset, it carries information
\emph{beyond} punctuation.

Punctuation words comprised 17.4\% of tokens (218/1256) and had higher
scores (means $-7.29$ vs.\ $-12.93$), with a point-biserial
correlation $r=0.604$ and AUROC (LLM score separating punctuation
vs.\ non-punctuation) $=0.858$, confirming a sensible association. As
a baseline predictor of human boundaries, \emph{punctuation alone}
achieved AUROC $=0.781$, AUPRC $=0.117$ against a prevalence baseline
$=0.039$, and $d'_{\text{AUC}}=1.10$; informative but markedly weaker
than the LLM score (AUROC $=0.945$, AUPRC $=0.428$,
$d'_{\text{AUC}}=2.26$). Crucially, the LLM score remained predictive
\emph{within} punctuation strata (AUROC $=0.929$ among non-punctuation
words; AUROC $=0.858$ among punctuation-only words) and after removing
the mean-level shift (residual AUROC $=0.917$). Thus, while
punctuation correlates with boundary likelihood, as expected, the
LLM-derived boundary metric captures additional event structure that
is not reducible to punctuation.

\clearpage

\section*{Supplementary Tables}

\begin{scriptsize}
\setlength{\tabcolsep}{5pt}%
\begin{longtable}{@{}>{\raggedright\arraybackslash}p{5.4cm} l r r@{}}
\caption{\textbf{Cross-story region of interest significance counts.} Number of
  stories (out of 13) in which region reached Simes $p{<}0.05$ for
  the \emph{drift} predictor ($\rho = 0.3$) and for the \emph{boundary}
  predictor. Hem = Hemisphere}\label{tab:roi_simes_combined}\\
\toprule
\textbf{ROI} & \textbf{Hem} & \textbf{Drift $\rho{=}0.3$} & \textbf{Boundary} \\
\midrule
\endfirsthead
\toprule
\textbf{ROI} & \textbf{Hem} & \textbf{Drift $\rho{=}0.3$} & \textbf{Boundary} \\
\midrule
\endhead
\midrule
\multicolumn{4}{r}{\emph{Continued on next page}}\\
\endfoot
\bottomrule
\endlastfoot
Planum Temporale & Left & 1 & 13 \\
Superior Temporal Gyrus (anterior) & Left & 1 & 13 \\
Heschl’s Gyrus (H1/H2) & Left & 0 & 13 \\
Planum Polare & Left & 0 & 13 \\
Superior Temporal Gyrus (anterior) & Right & 0 & 13 \\
Superior Temporal Gyrus (posterior) & Left & 0 & 13 \\
Superior Temporal Gyrus (posterior) & Right & 4 & 12 \\
Parietal Operculum Cortex & Left & 2 & 12 \\
Heschl’s Gyrus (H1/H2) & Right & 1 & 12 \\
Middle Temporal Gyrus (anterior) & Left & 1 & 11 \\
Planum Polare & Right & 0 & 11 \\
Central Opercular Cortex & Left & 1 & 10 \\
Middle Temporal Gyrus (anterior) & Right & 0 & 9 \\
Supramarginal Gyrus (posterior) & Left & 4 & 9 \\
Supramarginal Gyrus (posterior) & Right & 2 & 9 \\
Planum Temporale & Right & 1 & 9 \\
Middle Temporal Gyrus (posterior) & Right & 2 & 8 \\
Middle Temporal Gyrus (temporo-occipital) & Left & 3 & 8 \\
Middle Temporal Gyrus (posterior) & Left & 0 & 7 \\
Inferior Frontal Gyrus (pars triangularis) & Right & 2 & 7 \\
Supramarginal Gyrus (anterior) & Left & 4 & 4 \\
Middle Temporal Gyrus (temporo-occipital) & Right & 1 & 3 \\
Lateral Occipital Cortex (superior) & Left & 2 & 3 \\
Intracalcarine Cortex & Right & 1 & 3 \\
Supplementary Motor Area (Juxtapositional Lobule) & Left & 3 & 3 \\
Precuneus Cortex & Left & 4 & 3 \\
Precuneus Cortex & Right & 4 & 3 \\
Supracalcarine Cortex & Left & 2 & 3 \\
Middle Frontal Gyrus & Right & 2 & 3 \\
Precentral Gyrus & Left & 1 & 3 \\
Superior Parietal Lobule & Left & 4 & 2 \\
Superior Parietal Lobule & Right & 2 & 2 \\
Angular Gyrus & Right & 2 & 2 \\
Intracalcarine Cortex & Left & 2 & 2 \\
Cuneal Cortex & Left & 2 & 2 \\
Cuneal Cortex & Right & 1 & 2 \\
Occipital Fusiform Gyrus & Right & 1 & 2 \\
Central Opercular Cortex & Right & 1 & 2 \\
Parietal Operculum Cortex & Right & 3 & 2 \\
Supracalcarine Cortex & Right & 2 & 2 \\
Occipital Pole & Left & 0 & 2 \\
Occipital Pole & Right & 2 & 2 \\
Middle Frontal Gyrus & Left & 2 & 2 \\
Temporal Pole & Left & 0 & 2 \\
Temporal Pole & Right & 2 & 2 \\
Postcentral Gyrus & Left & 1 & 1 \\
Supramarginal Gyrus (anterior) & Right & 3 & 1 \\
Angular Gyrus & Left & 6 & 1 \\
Lateral Occipital Cortex (superior) & Right & 3 & 1 \\
Lateral Occipital Cortex (inferior) & Left & 2 & 1 \\
Medial Frontal Cortex & Right & 1 & 1 \\
Supplementary Motor Area (Juxtapositional Lobule) & Right & 2 & 1 \\
Posterior Cingulate Gyrus & Left & 1 & 1 \\
Lingual Gyrus & Left & 1 & 1 \\
Lingual Gyrus & Right & 1 & 1 \\
Superior Frontal Gyrus & Left & 2 & 1 \\
Occipital Fusiform Gyrus & Left & 4 & 1 \\
Inferior Frontal Gyrus (pars triangularis) & Left & 3 & 1 \\
Inferior Frontal Gyrus (pars opercularis) & Right & 3 & 1 \\
Precentral Gyrus & Right & 1 & 1 \\
Amygdala & Left & 0 & 0 \\
Accumbens & Left & 0 & 0 \\
Inferior Temporal Gyrus (anterior) & Left & 0 & 0 \\
Inferior Temporal Gyrus (anterior) & Right & 0 & 0 \\
Right Lateral Ventricle &  & 0 & 0 \\
Inferior Temporal Gyrus (posterior) & Left & 0 & 0 \\
Inferior Temporal Gyrus (posterior) & Right & 0 & 0 \\
Thalamus & Right & 0 & 0 \\
Inferior Temporal Gyrus (temporo-occipital) & Left & 0 & 0 \\
Inferior Temporal Gyrus (temporo-occipital) & Right & 1 & 0 \\
Caudate & Right & 0 & 0 \\
Postcentral Gyrus & Right & 2 & 0 \\
Putamen & Right & 1 & 0 \\
Hippocampus & Right & 0 & 0 \\
Frontal Pole & Left & 1 & 0 \\
Frontal Pole & Right & 1 & 0 \\
Amygdala & Right & 0 & 0 \\
Accumbens & Right & 0 & 0 \\
Lateral Occipital Cortex (inferior) & Right & 1 & 0 \\
Medial Frontal Cortex & Left & 1 & 0 \\
Subcallosal Cortex & Left & 3 & 0 \\
Subcallosal Cortex & Right & 3 & 0 \\
Paracingulate Gyrus & Left & 1 & 0 \\
Paracingulate Gyrus & Right & 1 & 0 \\
Anterior Cingulate Gyrus & Left & 1 & 0 \\
Anterior Cingulate Gyrus & Right & 1 & 0 \\
Insular Cortex & Left & 0 & 0 \\
Insular Cortex & Right & 1 & 0 \\
Posterior Cingulate Gyrus & Right & 4 & 0 \\
Cingulate Gyrus (posterior) & Right & 1 & 0 \\
Frontal Orbital Cortex & Left & 0 & 0 \\
Frontal Orbital Cortex & Right & 0 & 0 \\
Parahippocampal Gyrus (anterior) & Left & 0 & 0 \\
Parahippocampal Gyrus (anterior) & Right & 0 & 0 \\
Parahippocampal Gyrus (posterior) & Left & 2 & 0 \\
Parahippocampal Gyrus (posterior) & Right & 3 & 0 \\
Lingual Gyrus &  & 0 & 0 \\
Temporal Fusiform Cortex (anterior) & Left & 1 & 0 \\
Temporal Fusiform Cortex (anterior) & Right & 2 & 0 \\
Temporal Fusiform Cortex (posterior) & Left & 1 & 0 \\
Temporal Fusiform Cortex (posterior) & Right & 1 & 0 \\
Temporal Occipital Fusiform Cortex & Left & 1 & 0 \\
Temporal Occipital Fusiform Cortex & Right & 1 & 0 \\
Superior Frontal Gyrus & Right & 1 & 0 \\
Occipital Fusiform Gyrus & Right & 1 & 0 \\
Frontal Operculum Cortex & Left & 2 & 0 \\
Frontal Operculum Cortex & Right & 1 & 0 \\
Central Opercular Cortex & Left & 1 & 0 \\
Central Opercular Cortex & Right & 1 & 0 \\
Parietal Operculum Cortex & Left & 2 & 0 \\
Parietal Operculum Cortex & Right & 3 & 0 \\
Heschl’s Gyrus (H1/H2) & Right & 1 & 0 \\
Supracalcarine Cortex & Left & 2 & 0 \\
Supracalcarine Cortex & Right & 2 & 0 \\
Occipital Pole & Right & 2 & 0 \\
Thalamus & Left & 0 & 0 \\
Middle Frontal Gyrus & Left & 2 & 0 \\
Middle Frontal Gyrus & Right & 2 & 0 \\
Inferior Frontal Gyrus (pars triangularis) & Left & 2 & 0 \\
Caudate & Left & 0 & 0 \\
Inferior Frontal Gyrus (pars opercularis) & Left & 2 & 0 \\
Inferior Frontal Gyrus (pars opercularis) & Right & 3 & 0 \\
Putamen & Left & 1 & 0 \\
Pallidum & Left & 1 & 0 \\
Precentral Gyrus & Left & 1 & 0 \\
Precentral Gyrus & Right & 1 & 0 \\
Temporal Pole & Left & 0 & 0 \\
Temporal Pole & Right & 0 & 0 \\
Hippocampus & Left & 0 & 0 \\
Superior Temporal Gyrus (anterior) & Left & 1 & 0 \\
Superior Temporal Gyrus (anterior) & Right & 0 & 0 \\
\end{longtable}
\end{scriptsize}

\clearpage

\begin{scriptsize}
\setlength{\tabcolsep}{4.5pt} 
\begin{longtable}{@{}>{\raggedright\arraybackslash}p{4.75cm} l r r l r@{}}
\caption{\textbf{Region of interest-level directional contrast.} Mean
  $\overline{\Delta\beta}=\overline{\beta_{\text{shift}}-\beta_{\text{drift}}}$,
  Simes $p(\Delta\beta)$ across voxels, direction label
  (shift$\ge$drift / drift$\ge$shift / n.s.), and ROI size ($n$
  voxels). Full ROI names are used; hemisphere is listed
  separately. Hem = Hemisphere}
\label{tab:roi_unique_all}\\
\toprule
\textbf{ROI} & \textbf{Hem} & \textbf{$\overline{\Delta\beta}$} & \textbf{$p_{\text{Simes}}(\Delta\beta)$} & \textbf{Direction} & \textbf{$n$} \\
\midrule
\endfirsthead
\toprule
\textbf{ROI} & \textbf{Hem} & \textbf{$\overline{\Delta\beta}$} & \textbf{$p_{\text{Simes}}(\Delta\beta)$} & \textbf{Direction} & \textbf{$n$} \\
\midrule
\endhead
\midrule
\multicolumn{6}{r}{}\\
\endfoot
\bottomrule
\endlastfoot

Amygdala & Left & -0.000558 & 0.628836 & n.s. & 106 \\
Superior Temporal Gyrus (posterior division) & Left & 0.052598 & 4.7546e-07 & shift$\ge$drift & 392 \\
Superior Temporal Gyrus (posterior division) & Right & 0.046555 & 6.1714e-07 & shift$\ge$drift & 478 \\
Accumbens & Left & 0.000127 & 0.169430 & n.s. & 10 \\
Middle Temporal Gyrus (anterior division) & Left & 0.035144 & 4.2532e-06 & shift$\ge$drift & 165 \\
Middle Temporal Gyrus (anterior division) & Right & 0.017205 & 1.2101e-05 & shift$\ge$drift & 114 \\
Middle Temporal Gyrus (posterior division) & Left & 0.016087 & 2.9831e-05 & shift$\ge$drift & 365 \\
Middle Temporal Gyrus (posterior division) & Right & 0.021461 & 2.4058e-06 & shift$\ge$drift & 351 \\
Middle Temporal Gyrus (temporooccipital part) & Left & 0.011083 & 0.0054266 & shift$\ge$drift & 341 \\
Middle Temporal Gyrus (temporooccipital part) & Right & 0.001549 & 0.00021484 & shift$\ge$drift & 424 \\
Inferior Temporal Gyrus (anterior division) & Left & 0.005764 & 0.0189603 & shift$\ge$drift & 89 \\
Inferior Temporal Gyrus (anterior division) & Right & -0.000182 & 0.0074471 & drift$\ge$shift & 92 \\
Lateral Ventricle & Right & 0.000043 & 0.0917890 & n.s. & 272 \\
Inferior Temporal Gyrus (posterior division) & Left & 0.004657 & 0.0314186 & shift$\ge$drift & 327 \\
Inferior Temporal Gyrus (posterior division) & Right & 0.002930 & 0.0037946 & shift$\ge$drift & 325 \\
Thalamus & Right & 0.003165 & 0.385500 & n.s. & 80 \\
Inferior Temporal Gyrus (temporooccipital part) & Left & 0.007973 & 0.0187952 & shift$\ge$drift & 363 \\
Inferior Temporal Gyrus (temporooccipital part) & Right & -0.003653 & 0.0297292 & drift$\ge$shift & 342 \\
Caudate & Right & -0.002201 & 0.651655 & n.s. & 45 \\
Postcentral Gyrus & Left & -0.000610 & 0.0017277 & drift$\ge$shift & 1362 \\
Postcentral Gyrus & Right & -0.005481 & 0.0111283 & drift$\ge$shift & 1274 \\
Putamen & Right & -0.002019 & 0.0601635 & n.s. & 5 \\
Superior Parietal Lobule & Left & -0.007055 & 0.0217556 & drift$\ge$shift & 726 \\
Superior Parietal Lobule & Right & -0.004071 & 0.0677372 & n.s. & 452 \\
Hippocampus & Right & -0.003875 & 0.0126435 & drift$\ge$shift & 262 \\
Supramarginal Gyrus (anterior division) & Left & 0.010508 & 0.0191072 & shift$\ge$drift & 401 \\
Supramarginal Gyrus (anterior division) & Right & -0.001669 & 0.0137826 & drift$\ge$shift & 360 \\
Frontal Pole & Left & 0.002122 & 0.0993097 & n.s. & 3052 \\
Frontal Pole & Right & 0.004387 & 0.0212449 & shift$\ge$drift & 3841 \\
Amygdala & Right & -0.001834 & 0.188738 & n.s. & 149 \\
Supramarginal Gyrus (posterior division) & Left & 0.022588 & 5.1282e-07 & shift$\ge$drift & 433 \\
Supramarginal Gyrus (posterior division) & Right & 0.025459 & 9.0287e-06 & shift$\ge$drift & 725 \\
Angular Gyrus & Left & 0.009036 & 0.082200 & n.s. & 396 \\
Angular Gyrus & Right & 0.002227 & 4.8223e-05 & shift$\ge$drift & 577 \\
Accumbens & Right & 0.002391 & 0.0596555 & n.s. & 7 \\
Lateral Occipital Cortex (superior division) & Left & 0.003412 & 0.0170109 & shift$\ge$drift & 1929 \\
Lateral Occipital Cortex (superior division) & Right & -0.001109 & 0.00075845 & drift$\ge$shift & 1841 \\
Lateral Occipital Cortex (inferior division) & Left & -0.006172 & 0.0118370 & drift$\ge$shift & 858 \\
Lateral Occipital Cortex (inferior division) & Right & 0.000268 & 0.0225301 & shift$\ge$drift & 779 \\
Intracalcarine Cortex & Left & 0.010502 & 0.479807 & n.s. & 190 \\
Intracalcarine Cortex & Right & 0.010586 & 0.0305649 & shift$\ge$drift & 188 \\
Medial Frontal Cortex & Left & 0.003688 & 0.0330878 & shift$\ge$drift & 117 \\
Medial Frontal Cortex & Right & 0.015079 & 0.0180565 & shift$\ge$drift & 182 \\
Supplementary Motor Area (Juxtapositional Lobule) & Left & 0.006281 & 0.00020104 & shift$\ge$drift & 310 \\
Supplementary Motor Area (Juxtapositional Lobule) & Right & 0.002684 & 0.00012647 & shift$\ge$drift & 238 \\
Subcallosal Cortex & Left & 0.000831 & 0.0154952 & shift$\ge$drift & 290 \\
Subcallosal Cortex & Right & 0.001144 & 0.0053657 & shift$\ge$drift & 314 \\
Paracingulate Gyrus & Left & 0.003980 & 0.120625 & n.s. & 627 \\
Paracingulate Gyrus & Right & 0.001372 & 0.056431 & n.s. & 745 \\
Anterior Cingulate Gyrus & Left & -0.000990 & 0.0304911 & drift$\ge$shift & 452 \\
Anterior Cingulate Gyrus & Right & -0.001212 & 0.046760 & drift$\ge$shift & 518 \\
Insular Cortex & Left & 0.004344 & 0.0210726 & shift$\ge$drift & 579 \\
Insular Cortex & Right & 0.006991 & 0.0451745 & shift$\ge$drift & 507 \\
Posterior Cingulate Gyrus & Left & 0.001967 & 0.0397666 & shift$\ge$drift & 537 \\
Posterior Cingulate Gyrus & Right & 0.001918 & 0.0033638 & shift$\ge$drift & 499 \\
Precuneus Cortex & Left & 0.003812 & 0.0106011 & shift$\ge$drift & 1174 \\
Precuneus Cortex & Right & 0.002250 & 0.0222826 & shift$\ge$drift & 989 \\
Cuneal Cortex & Left & -0.001689 & 0.0298652 & drift$\ge$shift & 182 \\
Cuneal Cortex & Right & -0.001986 & 0.0309927 & drift$\ge$shift & 147 \\
Frontal Orbital Cortex & Left & 0.001413 & 0.0861123 & n.s. & 826 \\
Frontal Orbital Cortex & Right & 0.004478 & 0.0344690 & shift$\ge$drift & 717 \\
Parahippocampal Gyrus (anterior division) & Left & 0.002226 & 0.0767000 & n.s. & 578 \\
Parahippocampal Gyrus (anterior division) & Right & -0.004042 & 0.0026907 & drift$\ge$shift & 621 \\
Parahippocampal Gyrus (posterior division) & Left & 0.003034 & 0.0075103 & shift$\ge$drift & 328 \\
Parahippocampal Gyrus (posterior division) & Right & 0.000701 & 0.0116060 & shift$\ge$drift & 287 \\
Lingual Gyrus & Left & -0.000492 & 0.179905 & n.s. & 856 \\
Lingual Gyrus & Right & 0.004754 & 0.0090806 & shift$\ge$drift & 827 \\
Temporal Fusiform Cortex (anterior division) & Left & -0.000352 & 0.0808884 & n.s. & 178 \\
Temporal Fusiform Cortex (anterior division) & Right & -0.004111 & 0.0150039 & drift$\ge$shift & 157 \\
Temporal Fusiform Cortex (posterior division) & Left & 0.000391 & 0.0134519 & shift$\ge$drift & 455 \\
Temporal Fusiform Cortex (posterior division) & Right & -0.003932 & 0.142924 & n.s. & 332 \\
Temporal Occipital Fusiform Cortex & Left & 0.002801 & 0.0280758 & shift$\ge$drift & 430 \\
Temporal Occipital Fusiform Cortex & Right & 0.003651 & 0.0289416 & shift$\ge$drift & 518 \\
Superior Frontal Gyrus & Left & 0.005437 & 0.0554195 & n.s. & 1476 \\
Superior Frontal Gyrus & Right & 0.004250 & 0.00035849 & shift$\ge$drift & 1213 \\
Occipital Fusiform Gyrus & Left & -0.008533 & 0.0237896 & drift$\ge$shift & 597 \\
Occipital Fusiform Gyrus & Right & 0.009780 & 0.147971 & n.s. & 674 \\
Frontal Operculum Cortex & Left & 0.012396 & 0.0077966 & shift$\ge$drift & 213 \\
Frontal Operculum Cortex & Right & 0.008940 & 0.234855 & n.s. & 154 \\
Central Opercular Cortex & Left & 0.027159 & 5.6780e-06 & shift$\ge$drift & 401 \\
Central Opercular Cortex & Right & 0.021387 & 6.3127e-05 & shift$\ge$drift & 373 \\
Parietal Operculum Cortex & Left & 0.051082 & 7.9430e-06 & shift$\ge$drift & 190 \\
Parietal Operculum Cortex & Right & 0.009672 & 1.2898e-04 & shift$\ge$drift & 156 \\
Planum Polare & Left & 0.042650 & 7.7281e-06 & shift$\ge$drift & 147 \\
Planum Polare & Right & 0.040514 & 1.6406e-06 & shift$\ge$drift & 145 \\
Heschl’s Gyrus & Left & 0.095693 & 7.2479e-06 & shift$\ge$drift & 81 \\
Heschl’s Gyrus & Right & 0.068474 & 5.5663e-06 & shift$\ge$drift & 70 \\
Planum Temporale & Left & 0.091250 & 4.5656e-07 & shift$\ge$drift & 97 \\
Planum Temporale & Right & 0.041144 & 1.9731e-05 & shift$\ge$drift & 82 \\
Supracalcarine Cortex & Left & 0.009460 & 0.113525 & n.s. & 50 \\
Supracalcarine Cortex & Right & 0.019009 & 0.160583 & n.s. & 23 \\
Occipital Pole & Left & -0.003617 & 0.0511000 & n.s. & 835 \\
Occipital Pole & Right & 0.002611 & 0.109661 & n.s. & 748 \\
Thalamus & Left & 0.004953 & 0.100008 & n.s. & 123 \\
Middle Frontal Gyrus & Left & 0.004531 & 1.2366e-04 & shift$\ge$drift & 1180 \\
Middle Frontal Gyrus & Right & 0.005288 & 5.6822e-05 & shift$\ge$drift & 1334 \\
Inferior Frontal Gyrus (pars triangularis) & Left & 0.002106 & 0.0022778 & shift$\ge$drift & 365 \\
Inferior Frontal Gyrus (pars triangularis) & Right & 0.016964 & 0.0126339 & shift$\ge$drift & 314 \\
Caudate & Left & 0.004568 & 0.0398659 & shift$\ge$drift & 49 \\
Inferior Frontal Gyrus (pars opercularis) & Left & 0.003196 & 0.0100426 & shift$\ge$drift & 331 \\
Inferior Frontal Gyrus (pars opercularis) & Right & 0.006965 & 0.0020086 & shift$\ge$drift & 297 \\
Putamen & Left & -0.007069 & 0.110458 & n.s. & 5 \\
Pallidum & Left & -0.003162 & 0.115245 & n.s. & 3 \\
Precentral Gyrus & Left & 0.002591 & 5.9445e-05 & shift$\ge$drift & 1674 \\
Precentral Gyrus & Right & -0.001276 & 0.0132759 & drift$\ge$shift & 1481 \\
Temporal Pole & Left & 0.006104 & 2.4259e-05 & shift$\ge$drift & 1249 \\
Temporal Pole & Right & 0.006147 & 0.00022718 & shift$\ge$drift & 1264 \\
Hippocampus & Left & 0.002323 & 0.0053110 & shift$\ge$drift & 258 \\
Superior Temporal Gyrus (anterior division) & Left & 0.105594 & 3.1934e-07 & shift$\ge$drift & 128 \\
Superior Temporal Gyrus (anterior division) & Right & 0.078284 & 6.5573e-08 & shift$\ge$drift & 132 \\

\end{longtable}
\end{scriptsize}

\clearpage

\begingroup
\scriptsize
\setlength{\tabcolsep}{1pt}

\begin{longtable}{p{4.5cm} l l p{1.9cm} r r}
\caption{\textbf{Story corpus and scan schedule.} Narrated stories (German titles with canonical/original titles where applicable), authors, and scan schedule. Durations are computed from TR counts ($TR=1.18\,s$).}
\label{app:stories-schedule}\\

\toprule
\textbf{Canonical title} & \textbf{Author} & \textbf{Session/Run} & \textbf{Date and time} & \textbf{TRs} & \textbf{Minutes} \\
\midrule
\endfirsthead

\toprule
\textbf{Canonical title} & \textbf{Author} & \textbf{Session/Run} & \textbf{Date and time} & \textbf{TRs} & \textbf{Minutes} \\
\midrule
\endhead

\midrule
\endfoot

\bottomrule
\endlastfoot

\emph{La Main} & Guy de Maupassant & S01 / R1 & 2024-04-16\;16{:}20 & 686 & 13.5 \\
\emph{The Adventure of the Blue Carbuncle} & Arthur Conan Doyle & S02 / R1 & 2024-04-18\;15{:}18 & 2726 & 53.6 \\
\emph{Ligeia} & Edgar Allan Poe & S02 / R2 & 2024-04-18\;16{:}17 & 1781 & 35.0 \\
\emph{The Musgrave Ritual} & Arthur Conan Doyle & S03 / R1 & 2024-04-26\;15{:}05 & 2765 & 54.4 \\
\emph{King Pest} & Edgar Allan Poe & S03 / R2 & 2024-04-26\;16{:}03 & 1575 & 31.0 \\
\emph{The Five Orange Pips} & Arthur Conan Doyle & S04 / R1 & 2024-05-03\;13{:}57 & 2555 & 50.2 \\
\emph{The Premature Burial} & Edgar Allan Poe & S04 / R2 & 2024-05-03\;14{:}50 & 1904 & 37.4 \\
\emph{MS.\ Found in a Bottle} & Edgar Allan Poe & S06 / R1 & 2024-05-27\;14{:}57 & 1458 & 28.7 \\
\emph{The Black Cat} & Edgar Allan Poe & S06 / R2 & 2024-05-27\;15{:}28 & 1445 & 28.4 \\
\emph{The Sunningdale Mystery} & Agatha Christie & S06 / R3 & 2024-05-27\;15{:}58 & 1361 & 26.8 \\
\emph{The Stretelli Case} & Edgar Wallace & S07 / R1 & 2024-05-29\;14{:}26 & 1291 & 25.4 \\
\emph{The Masque of the Red Death} & Edgar Allan Poe & S07 / R2 & 2024-05-29\;14{:}54 & 833 & 16.4 \\
\emph{Die Pflanzen des Dr.\ Cinderella} & Gustav Meyrink & S07 / R3 & 2024-05-29\;15{:}12 & 1177 & 23.1 \\

\end{longtable}
\endgroup

\clearpage

\section*{Supplementary Figures}
\begin{figure*}[ht!]
  \centering
  \includegraphics[width=\linewidth]{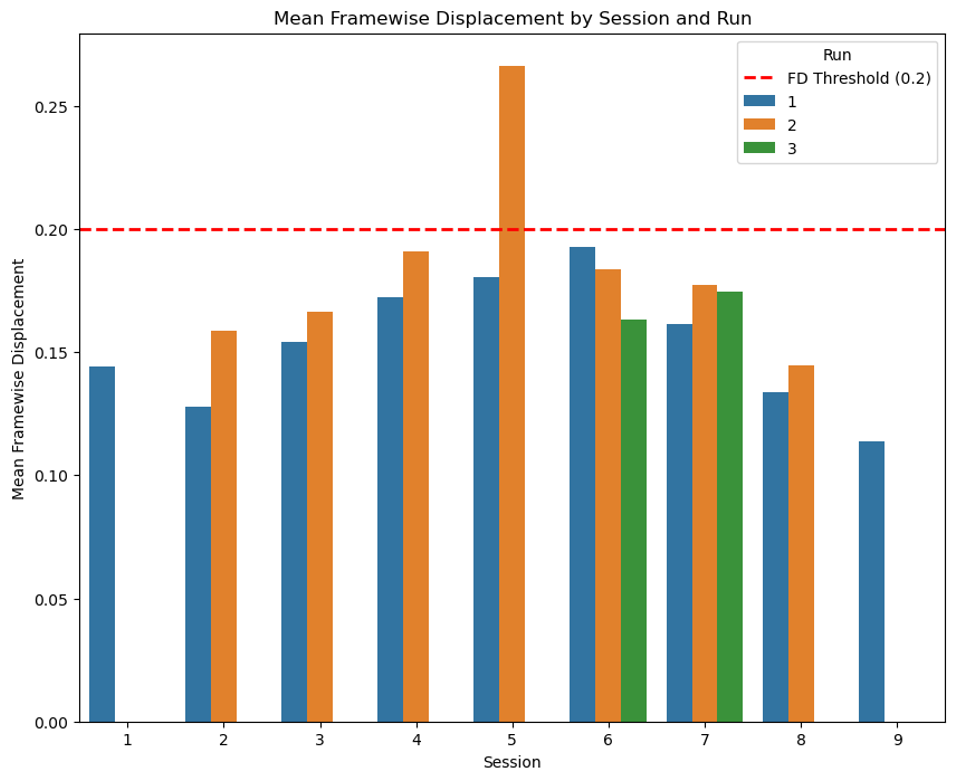}
  \caption{Mean frame-wise displacement in BOLD data.}
  \label{fig:app_md_displacement}
\end{figure*}

\clearpage

\begin{figure*}[ht!]
  \centering
  \includegraphics[width=\linewidth]{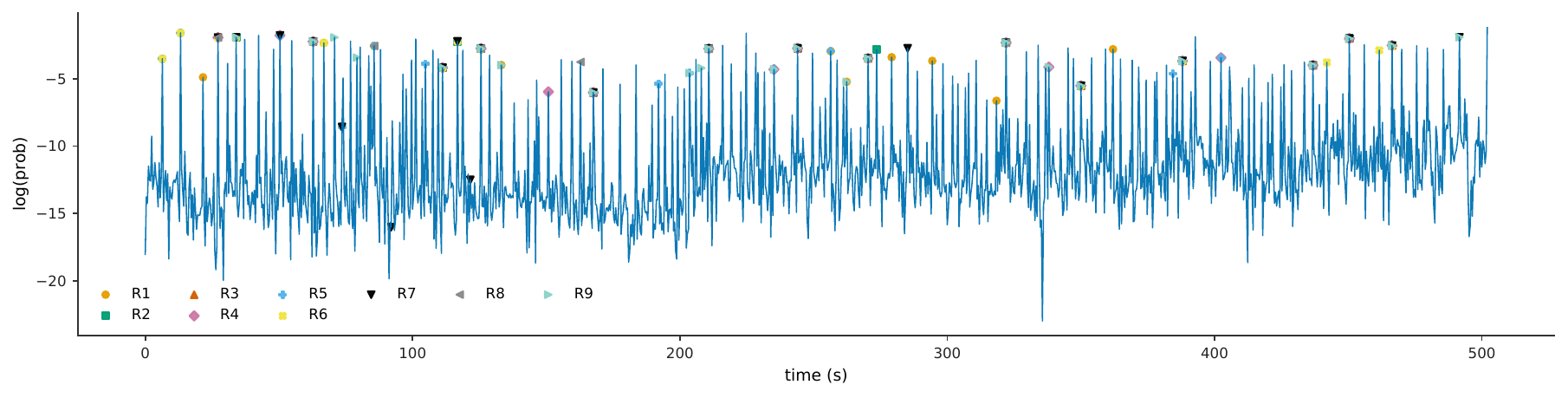}
  \caption{Boundary score over time (WORD level).  The curve shows the
    per-word \emph{log-probability} that the LLM would insert the
    boundary token ($\log p(\text{¶}\mid\text{context})$) aligned to
    word times. Colored markers indicate human event boundaries from
    individual raters (R1–R9), plotted at the word immediately
    preceding each annotated boundary.}
  \label{fig:app_timeseries}
\end{figure*}

\clearpage

\begin{figure*}[ht!]
  \centering
  \begin{subfigure}[t]{0.48\textwidth}
    \centering
    \includegraphics[width=\linewidth]{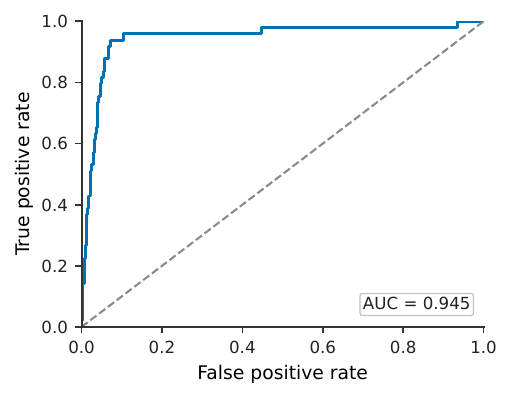}
    \subcaption{\textbf{ROC (AUROC = 0.945; 95\% CI 0.899–0.975).}}
  \end{subfigure}\hfill
  \begin{subfigure}[t]{0.48\textwidth}
    \centering
    \includegraphics[width=\linewidth]{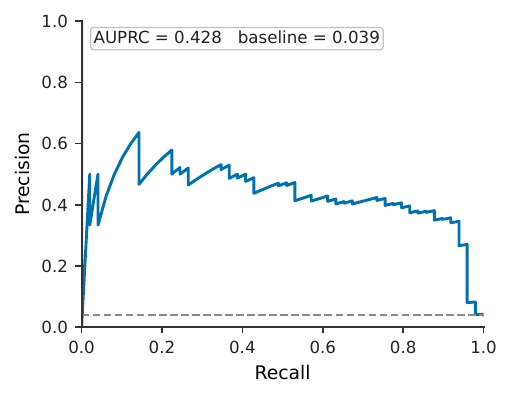}
    \subcaption{\textbf{Precision–Recall (AUPRC = 0.428; baseline = 0.039).}}
  \end{subfigure}
  \caption{\textbf{Threshold-free validation of the boundary metric.}
    (\textbf{Left}) ROC illustrates separability of boundary
    vs.\ non-boundary words (area under the curve, AUROC). Confidence
    intervals are block-bootstrapped with \SI{50}{word} blocks and
    $n_{\text{boot}}=10{,}000$ to respect temporal
    autocorrelation. (\textbf{Right}) PR curve shows performance under
    class imbalance; the dashed line marks the class-balance baseline
    (positives = 49/1{,}256 words). Together, these plots quantify
    discrimination without choosing a detection threshold.}
  \label{fig:app_rocpr}
\end{figure*}
\clearpage

\begin{figure*}[ht!]
  \centering
  \begin{subfigure}[t]{0.48\textwidth}
    \centering
    \includegraphics[width=\linewidth]{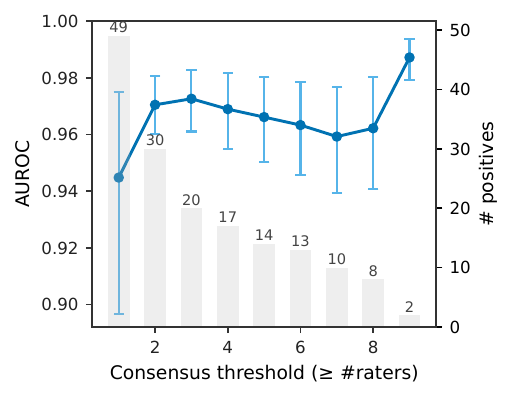}
    \subcaption{\textbf{Consensus sensitivity with 95\% CIs and \#positives.}}
  \end{subfigure}\hfill
  \begin{subfigure}[t]{0.48\textwidth}
    \centering
    \includegraphics[width=\linewidth]{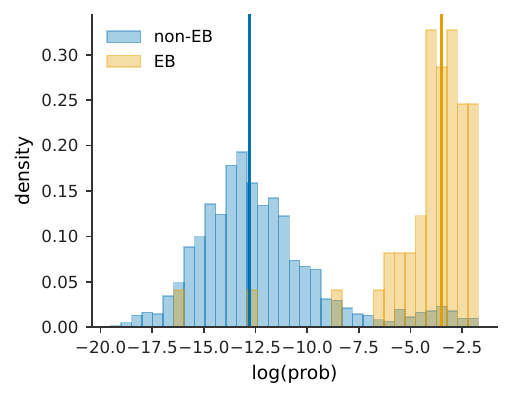}
    \subcaption{\textbf{Score distributions for EB vs.\ non-EB words.}}
  \end{subfigure}
  \caption{\textbf{Calibration to human agreement and score
      separation.}  (\textbf{Left}) AUROC as the consensus threshold
    increases ($\ge$\#raters). Shaded bars (right axis) give the
    number of positives at each threshold, clarifying uncertainty when
    few positives remain. The metric stays highly discriminative
    across mid-to-high consensus levels. (\textbf{Right}) Histograms
    of the boundary score (log-prob) for positive vs.\ negative words
    with median markers; the parametric SDT effect size across the two
    distributions is $d'=2.85$.}
  \label{fig:app_consensus_dists}
\end{figure*}
\clearpage

\begin{figure*}[ht!]
  \centering
  \begin{subfigure}[H]{0.49\textwidth}
    \centering
    \includegraphics[width=\linewidth]{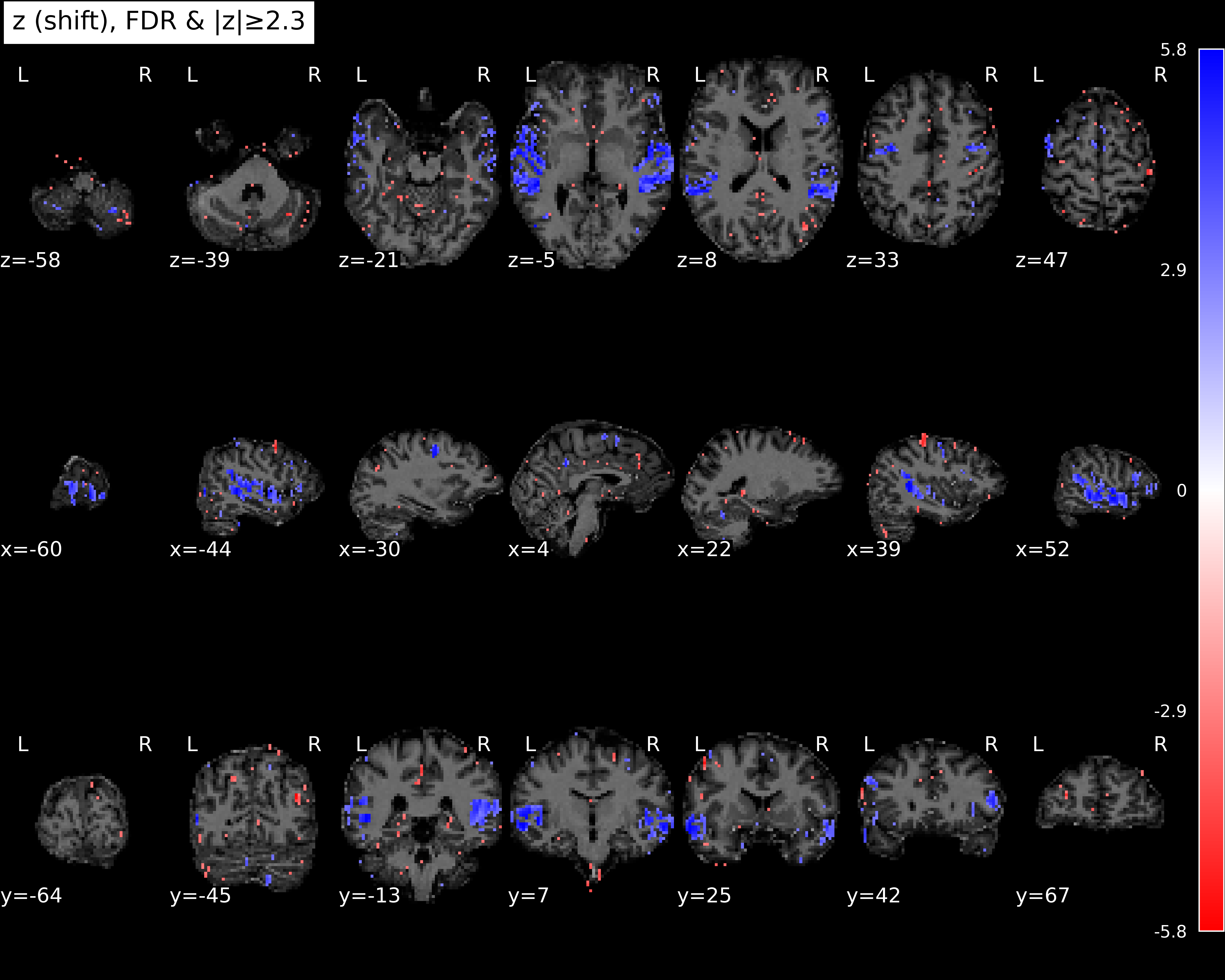}
    \subcaption{\textbf{Unique shift (vs.\ punctuation)}}
  \end{subfigure}\hfill
  \begin{subfigure}[H]{0.49\textwidth}
    \centering
    \includegraphics[width=\linewidth]{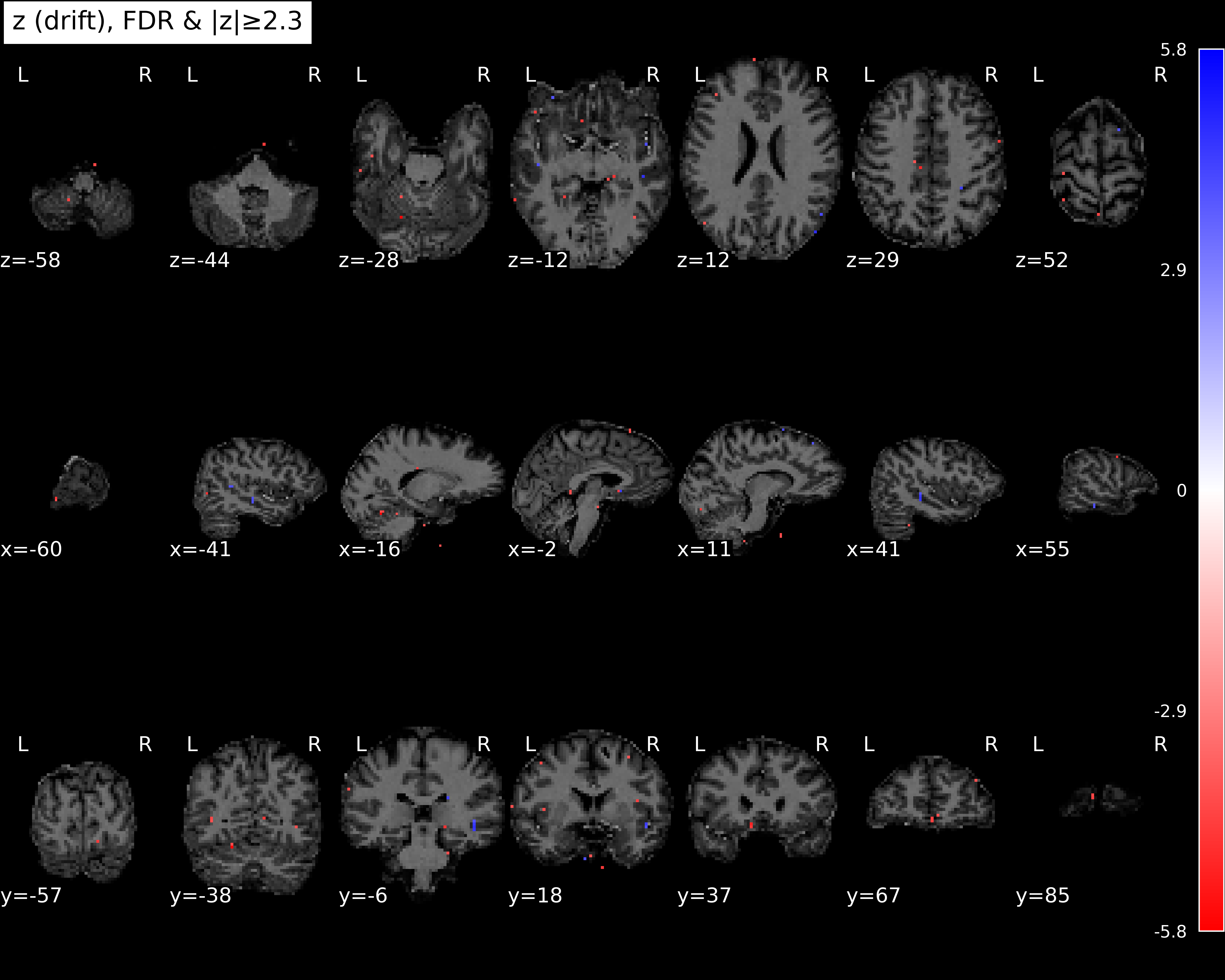}
    \subcaption{\textbf{Unique punctuation (vs.\ shift)}}
  \end{subfigure}
  \caption{\textbf{Punctuation control: voxelwise unique-effects
      maps.}  Across-story one-sample $t{\to}z$ maps (df$=12$) for the
    pairwise stage-2 model $y(t)=\beta_{\text{shift}}\hat
    y_{\text{shift}}(t)+\beta_{\text{punct}}\hat
    y_{\text{punct}}(t)+\varepsilon(t)$.  Maps are BH–FDR masked
    ($q{=}0.05$) with a display threshold of $|z|\ge 2.3$ under a
    symmetric diverging colormap.  \textbf{Unique shift} shows robust,
    bilateral clusters along the peri-Sylvian belt (Heschl's gyrus,
    planum temporale/polare, STG/MTG), whereas \textbf{unique
      punctuation} is markedly weaker and sparser in these same
    regions, indicating that boundary-linked responses are not
    reducible to pauses/typography.}
  \label{fig:app_punct_unique_zmaps}
\end{figure*}

\end{document}